\documentclass[prx,twocolumn,superscriptaddress,longbibliography,footinbib]{revtex4-2}

\usepackage[utf8]{inputenc}

\usepackage{graphicx}
\graphicspath{{./figures/}}

\usepackage[dvipsnames]{xcolor}
\definecolor{dark-blue}{rgb}{0,0.2,0.6}

\usepackage{etoolbox}
\makeatletter
\pretocmd{\NAT@open}{\begingroup\color{\@citecolor}}{}{}
\apptocmd{\NAT@close}{\endgroup}{}{}
\makeatother

\usepackage{stix}
\DeclareMathAlphabet{\mathcal}{OMS}{cmsy}{m}{n}
\DeclareSymbolFont{CMAlt}{OMX}{cmex}{m}{n}
\DeclareMathSymbol{\sumop}{\mathop}{CMAlt}{"50}
\DeclareMathSymbol{\intop}{\mathop}{CMAlt}{"52}

\usepackage{amsmath}
\usepackage{mathtools}

\usepackage{upgreek}

\usepackage{xspace}

\usepackage[colorlinks,%
            pdfusetitle,%
            urlcolor=dark-blue,%
            citecolor=dark-blue,%
            linkcolor=dark-blue]{hyperref}

\newcommand{\ket}[1]{\ensuremath{\left|{#1}\right\rangle}}
\newcommand{\fixedket}[1]{\ensuremath{|{#1}\rangle}}

\newcommand{\pFq}[2]{\ensuremath{{}_{#1}F_{#2}}}

\DeclarePairedDelimiter\abs{\lvert}{\rvert}

\newcommand{\be}{\begin{equation}}
\newcommand{\ee}{\end{equation}}

\newcommand{\yb}{\ensuremath{{^\text{174}\text{Yb}}}\xspace}
\newcommand{\ybf}{\ensuremath{{^\text{171}\text{Yb}}}\xspace}

\newcommand{\tP}[1]{\ensuremath{{^3\text{P}_{#1}}}}
\newcommand{\tD}[1]{\ensuremath{{^3\text{D}_{#1}}}}

\newcommand{\sS}[1]{\ensuremath{{^1\text{S}_{#1}}}}

\newcommand{\sP}[1]{\ensuremath{{^1\text{P}_{#1}}}}

\newcommand{\nbar}[1]{\ensuremath{{\bar{n}_{#1}}}}
\newcommand{\gsf}[1]{\ensuremath{{\mathcal{F}_{#1\text{D}}}}}

\newcommand{\mathunit}[1]{\ensuremath{\,#1}}
\newcommand{\asciimathunit}[1]{\ensuremath{\,\text{#1}}}

\newcommand{\um}{\mathunit{\upmu\text{m}}}
\newcommand{\nm}{\asciimathunit{nm}}

\newcommand{\ms}{\asciimathunit{ms}}
\newcommand{\us}{\mathunit{\upmu\text{s}}}

\newcommand{\MHz}{\asciimathunit{MHz}}
\newcommand{\kHz}{\asciimathunit{kHz}}

\newcommand{\uK}{\mathunit{\upmu\text{K}}}
\newcommand{\nK}{\asciimathunit{nK}}

\newcommand{\mW}{\asciimathunit{mW}}

\newcommand{\Gauss}{\asciimathunit{G}}

\newcommand{\Erec}{E_\text{rec}}
\newcommand{\rec}{\omega_\text{rec}}

\newcommand{\subfigref}[2]{\hyperref[fig:#1]{\ref*{fig:#1}(#2)}}
\newcommand{\panel}[1]{$\left(\text{#1}\right)$}

\begin{document}


\title{Isotope-agnostic motional ground-state cooling of neutral Yb atoms}

\author{Ronen~M.~Kroeze}
\author{Ren\'e~A.~Villela}
\author{Er~Zu}
\author{Tim~O.~H\"ohn}

\affiliation{Fakult{\"a}t f{\"u}r Physik, Ludwig-Maximilians-Universit{\"a}t, 80799 M{\"u}nchen, Germany}
\affiliation{Munich Center for Quantum Science and Technology (MCQST), 80799 M{\"u}nchen, Germany}

\author{Monika~Aidelsburger}
\email{Monika.Aidelsburger@mpq.mpg.de}

\affiliation{Fakult{\"a}t f{\"u}r Physik, Ludwig-Maximilians-Universit{\"a}t, 80799 M{\"u}nchen, Germany}
\affiliation{Munich Center for Quantum Science and Technology (MCQST), 80799 M{\"u}nchen, Germany}
\affiliation{Max-Planck-Institut f{\"u}r Quantenoptik, 85748 Garching, Germany}


\begin{abstract}
Efficient high-fidelity ground-state cooling of motional degrees of freedom is crucial for applications in quantum simulation, computing and metrology.
Here, we demonstrate direct ground-state cooling of fermionic \ybf and bosonic \yb atoms in two- and three-dimensional magic-wavelength optical lattices on the ultranarrow clock transition.
Its high spectral resolution offers the potential for reaching extremely low temperatures.
To ensure efficient cooling, we develop a chirped sideband cooling scheme, where we sweep the clock-laser frequency to mitigate the effects of spatial trap inhomogeneities.
We further generalize the theoretical description of sideband spectra to higher-dimensional lattices for precise thermometry.
We achieve 2D ground state fractions of $97\%$ for \ybf with an average motional occupation of $\bar{n}\simeq0.015$ and provide a direct comparison with \yb, reaching similar cooling performance.
Applying the same scheme in 3D results in $\bar{n}\simeq0.15$ limited by layer-to-layer inhomogeneities in the vertical direction. 
These results demonstrate efficient motional ground-state cooling in optical lattices, especially for bosonic alkaline-earth(-like) atoms, where other methods are not applicable, opening the door to novel protocols for quantum science applications with neutral atoms.
\end{abstract}

\maketitle


Neutral atoms in optical arrays play a central role in quantum science, from quantum simulation of itinerant physics~\cite{Gross_Quantum_2017,gross_quantum_2021} to optical clocks~\cite{Ludlow_Optical_2015,ushijima_cryogenic_2015,mcgrew_atomic_2018,Aeppli_Clock_2024,li_strontium_2024}, quantum computation~\cite{Kaufman_Quantum_2021,henriet_quantum_2020} and entanglement-enhanced metrology~\cite{schine_long-lived_2022,cao_multi-qubit_2024,robinson_direct_2024,zaporski_quantum-amplified_2025}.
To achieve optimal performance, ground-state cooling is essential.
In digital quantum computing, motional excitations severely limit gate fidelities~\cite{robicheaux_photon-recoil_2021,schine_long-lived_2022,Lis_Midcircuit_2023}, 
while in optical lattice clocks, finite temperatures induce systematic uncertainties~\cite{Brown_Hyperpolarizability_2017}.
Although experimental protocols based on spilling or removal of motional excited states have been developed~\cite{serwane_deterministic_2011,young_tweezer-programmable_2022,Aeppli_Clock_2024,Shaw_Erasure_2025}, fast low-loss cooling to the absolute motional ground state remains a significant challenge for quantum technologies.

Alkaline-earth(-like) atoms (AELA) are particularly suitable for quantum applications given their favorable internal level structure with different narrow and broad transitions for efficient cooling, imaging and internal-state manipulation. 
Several techniques have been developed for achieving sub-recoil cooling~\cite{Schreck_Laser_2021,phatak_generalized_2024}.
In particular, Raman-sideband cooling, originally developed for alkali atoms, has proven effective both in tweezer arrays~\cite{Kaufman_Cooling_2012, Jenkins_Ytterbium_2022} and optical lattices~\cite{Hamann_Resolved-Sideband_1998,kerman_beyond_2000,Norcia_Iterative_2024,shu_increased_2024}.
However, this scheme inherently relies on a ground-state hyperfine structure, which is only available for fermionic AELA isotopes, having nuclear spin $I > 0$.

\begin{figure}[t!]
	\includegraphics[width=\columnwidth]{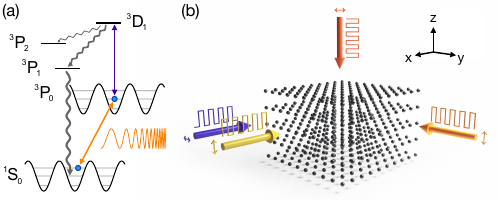}
	\caption{\label{fig:1}
		\textbf{Schematic of the clock-sideband cooling scheme}.
		\panel{a} Basic internal level structure with optical clock $\sS{0}\leftrightarrow\tP{0}$ and repumping $\tP{0}\leftrightarrow\tD{1}$ transition.
        Clock-sideband transitions are driven by multiple pulses while sweeping the frequency. 
        The wavy arrows indicate spontaneous decay paths.
        \panel{b} Illustration of the setup: 3D optical lattice with three independent pulsed clock-laser beams (yellow, orange, red) and one repumping beam (purple).
}
\end{figure}

Alternatively, direct resolved sideband cooling has been demonstrated for Sr atoms on the $\sS{0}\leftrightarrow\tP{1}$ transition, which uniquely features a narrow linewidth of a few kilohertz combining high-spectral resolution with large scattering rates~\cite{Norcia_Microscopic_2018, Cooper_Alkaline-Earth_2018}.
In Yb the equivalent transition is rather broad ($2\pi \times 183\kHz$) and direct sideband cooling is instead realized on the ultranarrow optical clock transition [Fig.~\subfigref{1}{a}].
In particular, 1D clock sideband cooling (CSC) is a standard technique in fermionic optical lattice clocks to reach record-low temperatures~\cite{Nemitz_Frequency_2016}.
Recently, this has been combined with clock-assisted cooling methods in the weakly confined radial direction to further enhance clock performance~\cite{Zhang_Subrecoil_2022,Chen_Clock-Line-Mediated_2024}.
Here we extend CSC to higher-dimensional 2D and 3D lattices and specifically demonstrate its applicability for bosonic \yb where other sub-Doppler cooling methods fail.
Moreover, we extend the theoretical model of sideband spectroscopy beyond standard 1D lattices~\cite{Blatt_Rabi_2009} by incorporating potential inhomogeneities, which is essential for benchmarking the cooling performance and achieved final temperatures of the chirped CSC scheme.
Our results pave the way for applications in quantum science with optical tweezer arrays and lattices, which require a high absolute ground-state fraction.

All experiments start by loading a laser-cooled cloud of Yb atoms from a magneto-optical trap on the $\sS{0}\leftrightarrow\tP{1}$ transition at $556\,$nm into a clock-magic optical lattice at $\lambda=759\nm$~\cite{Barber_Optical_2008}.
The lattice is generated by two retro-reflected laser beams along $x$ and $y$ and a vertical interfering lattice with $2.2\um$ spacing between layers along $z$ [Fig.~\subfigref{1}{b}].
To isolate atoms in the crossing region of the lattices and remove atoms trapped in the wings, we first load the atoms into a deep 3D lattice, then turn off the horizontal ones while increasing the vertical confinement within $10\ms$, followed by $10\ms$ equilibration time before ramping the lattices to the final trap configuration.
To perform sideband cooling we employ three independent clock-laser beams at $\lambda_c=578\,$nm that are approximately aligned with each lattice axis and drive the red sideband connecting $|\sS{0},n \rangle \rightarrow |\tP{0},n-1\rangle$.
To close the cooling cycle we apply a laser beam resonant with the $\tP{0}\leftrightarrow \tD{1}$ transition at $1388\nm$ along $x$, which after cascaded spontaneous decay via the \tP{1} state returns to \sS{0}, ideally preserving the motional quantum number.
A small fraction of atoms (2.5$\%$) decays to $\tP{2}$, which is anti-trapped and results in atom loss.
We opt to use pulsed rather than continuous cooling which avoids differential light shifts on the clock transition due to the resonant repumper.
Finally we detect the atoms via fluorescence imaging with molasses cooling on the $\sS{0} \leftrightarrow \tP{1}$ transition in a deep 3D lattice~\cite{Hohn_Determining_2024}.

\begin{figure}[t!]
	\includegraphics{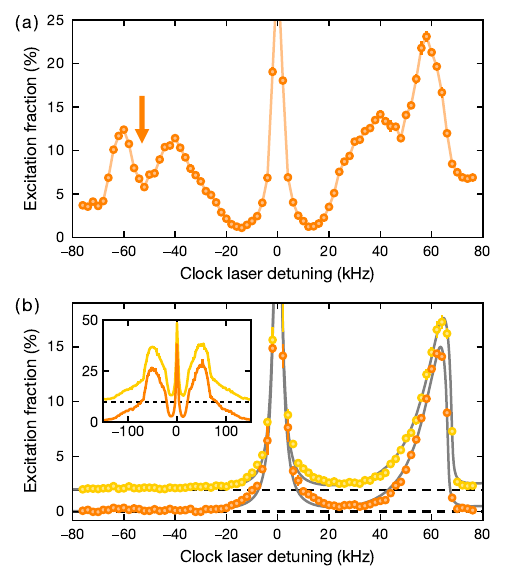}
	\caption{\label{fig:2}
		\textbf{Sideband cooling in a 2D lattice, illustrated for \ybf}.
		\panel{a} Inefficacy of cooling with constant clock-laser frequency.
        Spectroscopy along $y$ reveals a non-thermal red sideband, showing a dip in the vicinity of the cooling frequency (indicated by the arrow).
        The solid line connects the data points and is a guide to the eye.
        In both panels error bars represent the standard error over three repetitions and typically are smaller than the symbols~\cite{SM}.
        \panel{b} Sideband spectrum after chirped CSC.
        Spectra along $x$ (yellow) and $y$ (orange) are taken after 15 cooling cycles, the former is vertically offset (dashed lines indicate zero) for better visibility.
        The gray solid lines are fits with the model discussed in the main text, extracting $\nbar{x}=0.013^{+2}_{-4}$ and $\nbar{y}=0.018^{+2}_{-3}$; confidence intervals are evaluated with bootstrapping.
        Inset: spectra of initial state before cooling.
}
\end{figure}

Loading a laser-cooled cloud of atoms directly into a 2D or 3D lattice configuration typically results in system sizes that are comparable with the waist of the lattice laser beams.
While this is beneficial for fast loading schemes, the atoms experience inhomogeneous trap frequencies challenging efficient cooling.
This effect is notably more pronounced than in 1D lattices, as discussed below, and renders cooling at a single, constant sideband frequency ineffective even in magic trapping potentials [Fig.~\subfigref{2}{a}], as used in this work.
To overcome this challenge, we employ a chirped CSC scheme, where the frequency of the sideband cooling beam is slowly swept following a linear frequency ramp during the pulsed cooling sequence introduced above.
Similar schemes have been developed previously for non-magic trapping potentials~\cite{Berto_Prospects_2021, Holzl_Motional_2023}.

We start by demonstrating 2D chirped CSC of fermionic $^{171}$Yb atoms in a deep 2D horizontal lattice with depth $V_x=V_y\simeq 300\Erec$ and Lamb-Dicke parameter $\eta_x=\eta_y\simeq0.22$, where $\Erec=h^2/(2m\lambda^2)$ is the recoil energy, $\eta_i=\sqrt{h\pi/(m\omega_i \lambda_c^2)}$, $\omega_i$ is the harmonic trap frequency, $h$ is Planck's constant and $m$ is the mass of Yb.
CSC is applied along the two horizontal axes using a calibrated carrier Rabi frequency of $\Omega/(2\pi)=3.7\kHz$.
We want to emphasize that for \ybf much higher clock Rabi frequencies could be realized, but we employ this modest strength to facilitate direct comparison to cooling of bosonic \yb, where a magnetic field is required to induce a coupling between the two states~\cite{Taichenachev_Magnetic_2006}.
We find robust ground-state cooling for pulse durations of $500\us$, which are interleaved with $250\us$-long repumping pulses.
To address all atoms efficiently we apply a linear ramp of the clock-laser detuning from $-40\kHz$ to $-72\kHz$.
Each sweep consists of 20 alternating cooling and repumping pulses.
To cool both axes we further alternate between both directions.
The combination of one sweep along each axis defines one complete ``cooling cycle''.

After loading the laser-cooled atoms directly into the lattice the average occupation of excited states is rather high with $\nbar{x,y}\approx6$ [inset Fig.~\subfigref{2}{b}].
This is measured using sideband spectroscopy on the clock transition with a carrier Rabi frequency of $\Omega'/(2\pi)=1.1\kHz$ and a pulse duration of $20\ms$.
Repeating this sideband spectroscopy after 15 cooling cycles we obtain a high 2D motional ground state fraction of $\gsf{2}=0.970^{+5}_{-2}$ [Fig.~\subfigref{2}{b}] with nearly identical cooling performance along both axes reaching ultra-low excited state populations with $\nbar{}\simeq0.015$, where $\gsf{2} \approx (1+\nbar{x})^{-1}(1+\nbar{y})^{-1}$.

Such detailed quantitative analysis of the high-resolution clock sideband spectra crucially relies on an accurate modeling of the trap frequency inhomogeneities and thermal excitations.
A rigorous mathematical derivation is presented in the Supplementary Material~\cite{SM} and we briefly summarize the main aspects here.
For the coldest temperatures, the sideband spectrum $P(\tilde{\omega})$ consists of the carrier and the two first-order sidebands and can be expressed in the following form
\be\begin{split}\label{eq:spectrum}
	P(\tilde{\omega})=& A_0 L_{\Gamma_0}(\tilde{\omega}) \\
	& + A_1 L_{\Gamma_1}(\tilde{\omega})* \left[\rho_{(0)}(\tilde{\omega}) + \rho_{(1)}(-\tilde{\omega}-\rec)\right],
\end{split}\ee
where $A_{0,1}$ denote the amplitudes of the carrier and first-order sidebands, $\tilde{\omega}$ denotes the laser frequency detuning from the free-space clock resonance, $\hbar\rec=\Erec$ is the lattice recoil energy and we explicitly introduce the kernel $\rho(\tilde{\omega})$, whose convolution with a Lorentzian $L_\Gamma$ determines the shape of the sideband, where we allow the linewidth $\Gamma$ to be different for the carrier ($\Gamma_0$) and the first-order sidebands ($\Gamma_1$); $\rho_{(n_0)}$ is a modified sideband kernel defined below, which accounts for the fact that there are no $n_0$-th order red sideband transitions for the lowest $n_0$ motional states.

Let us start by reviewing the result for 1D lattices, where atoms are strongly confined along $x$, but weakly confined along $y$ and $z$.
Semi-classically, radial motion results in atoms sampling regions of the trapping potential where the longitudinal trap frequency $\omega_x$ is reduced.
Hence, the sideband shape depends on the radial temperature $T_r$ resulting in a characteristic sideband given by~\cite{Blatt_Rabi_2009, SM}
\begin{eqnarray}
	\rho_{(n_0)}(\tilde{\omega}) &=& Z ^{-1}\sum\limits_{n=n_0} e^{-\frac{n\hbar\omega_x}{k_BT_x}}f_{1\text{D}}\left(\omega_x - \tilde{\omega} - \rec(n+1)\right), \label{eq:sidebandKernel} \\
    f_{1\text{D}}(s) &=& s e^{-\alpha s}\Theta(s),\label{eq:shape1d}
\end{eqnarray}
where $Z$ is the partition function, $\alpha = \hbar\omega_x/(\rec k_BT_r)$, and $\Theta(s)$ is the Heaviside step function.
Notably, $\rho$ depends on the radial temperature through $\alpha$ but not the radial confinement or the waist of the lattice beam.
The result is a sideband spectrum where the relative height of the red versus blue sidebands is determined by the longitudinal temperature $T_x$, while the width of the sideband encodes the radial temperature $T_r$.

In a 2D lattice, the atoms are confined in an array of 1D tubes with strong confinement along $x$ and $y$, but weak confinement along $z$.
Similar to 1D, the atoms sample regions of the weakly-confined $z$ axis, resulting in a dependence on $T_z$ but not $T_y$, where the atoms are strongly confined.
This leads to a modified sideband kernel with
\be\label{eq:shape2d}
	f_{2\text{D}}(s) = e^{-\beta s}\Theta(s),
\ee
and $\beta = \hbar\omega_x(1+\omega_y^2/\omega_x^2)/(\rec k_BT_z)$.
Note that the linear prefactor $s$, which accounted for the radial symmetry in 1D, disappeared, which results in a more sharply-peaked sideband.
This result holds for large homogeneous 2D lattices.
However, if the trap frequency varies between tubes over the extent of the atomic cloud, this expression needs to be further averaged over spatial inhomogeneities and the shape of the sideband spectra explicitly depends on the relative size of the atomic cloud.
In this work the waist of the lattice laser beams is $w_0\simeq40\um$ and the size of the atomic cloud is $\sigma\simeq12\um$ (defined as the Gaussian standard deviation), hence, inhomogeneous broadening of the sideband spectra is significant.
While the relative height of the red and blue sideband still reveals information about the horizontal temperature $T_x$, or $T_y$, the width of the sideband now depends on the vertical temperature $T_z$ and the inhomogeneous broadening.
This is in stark contrast to 1D lattices, where layer-to-layer variations are set by the Rayleigh range and can typically be safely ignored.

In a 3D lattice atoms are strongly confined along all axes and the atoms cannot sample regions with lower trap frequency.
The new kernel is now given by the simple Dirac $\delta$-function $f_{3\text{D}}(s) = \delta(s)$.
In this case the shape of the sideband is mostly dominated by spatial inhomogeneities.
Assuming an isotropic Gaussian population with standard deviation $\sigma$ and radially symmetric lattice beams with waist $w_0$, the site-averaged sideband kernel is still given by Eq.~\eqref{eq:sidebandKernel} but with $f_{1\text{D}}(s)$ replaced by
\be
    \overline{f_{3\text{D}}}(s) = \frac{\gamma}{\omega_x}\left(1 - \frac{s}{\omega_x}\right)^{\gamma-1}\Theta(\omega_x-s)\Theta(s),
\ee
where $\gamma = w_0^2/(2\sigma^2)$.
Thus, the sideband now acquires a polynomially peaked shape, where the width encodes information about the spatial distribution.

\begin{figure}[t!]
	\includegraphics{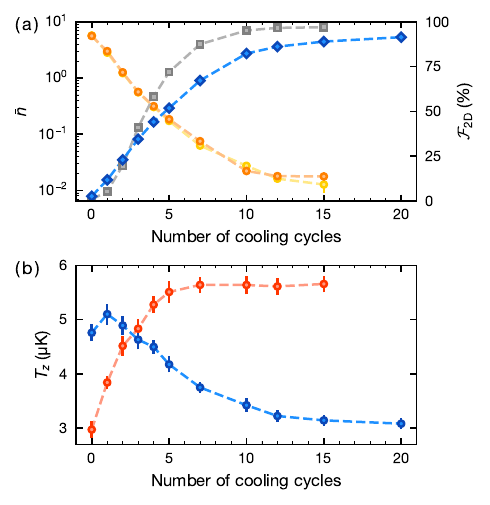}
	\caption{\label{fig:3}
		\textbf{Comparison of cooling performance of 2D CSC for \ybf and \yb}.
		\panel{a} Mean motional occupation number $\nbar{}$ (left axis) versus number of cooling cycles for \ybf, measured with sideband spectroscopy along $x$ (yellow) and $y$ (orange).
        The reduction of $\nbar{}$ coincides with an increase of the 2D motional ground state fraction $\gsf{2}$ (right axis), reaching $0.970^{+5}_{-2}$ after 15 cooling cycles for \ybf (gray squares) and $\gsf{2}=0.915^{+5}_{-5}$ after 20 cooling cycles for \yb (blue diamonds).
        Error bars from bootstrapping are all smaller than the data points.
        \panel{b} Time-of-flight thermometry of the out-of-plane, vertical temperature for \ybf (red) and \yb (blue).
        Error bars indicate the uncertainty in fitting the time-of-flight expansion.
}
\end{figure}

Given the theoretical expressions defined in Eqs.~\eqref{eq:spectrum}, \eqref{eq:sidebandKernel}, and \eqref{eq:shape2d}, we extract the corresponding temperatures from the measured sideband spectra [Fig.~\subfigref{2}{b}], where the lattice depth and amplitudes $A_{0,1}$ are additional free fit parameters.
From these fits, we track the cooling performance as a function of the number of applied cooling cycles along both axes [Fig.~\subfigref{3}{a}].
Note that after direct loading into the lattice, atoms approximately equally populate all $\simeq\!11$ bands of the lattice.
This extreme regime is not accurately captured by our model, which otherwise matches the measured spectra well once the temperature is lower.

\begin{figure*}[t!]
	\includegraphics{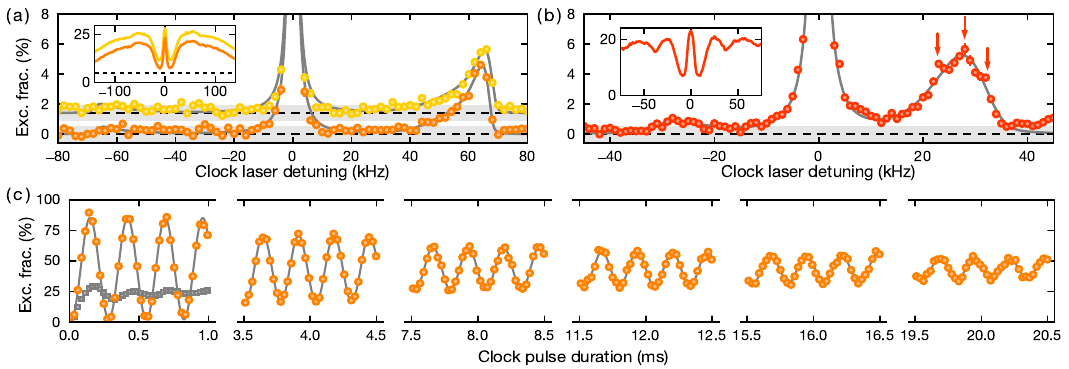}
	\caption{\label{fig:4}
		\textbf{3D clock cooling of \ybf}.
		\panel{a,b} Sideband spectra of the horizontal directions \panel{a} and vertical direction \panel{b}, after 3D cooling.
        Insets show spectra prior to cooling.
        Gray shaded area indicates standard deviation of background noise present in imaging without any atoms present.
        The solid line is a fit where the sideband has a generic asymmetric lineshape, see text for details.
        From the sideband asymmetry we extract $(\nbar{x},\nbar{y},\nbar{z})\simeq(0.17, 0.12, 0.16)$, corresponding to a 3D groundstate fraction of $66\%$.
        Data is averaged over ten repetitions [three for panel \panel{c}], their standard error is typically smaller than the point.
		\panel{b} The vertical spectrum shows resolvable peaks in the blue sideband, indicated by arrows, corresponding to the different layers.
		\panel{c} Rabi oscillations prior to cooling (gray squares) versus after 3D cooling (orange circles), driven by one of the horizontal beams.
        The former rapidly dephases due to the thermal mixture of Rabi frequencies, the latter shows persistent oscillation.
        The solid gray line is a separate fit for each segment, the contrast decays approximately exponentially with $1/e$ time of $11.4(2)\ms$.}
\end{figure*}

Since CSC does not require any internal structure, it is applicable to all isotopes.
To demonstrate this, we implement chirped CSC of bosonic \yb in a 2D lattice with equal depth and identical cooling sequence.
For \yb, the clock transition is magnetically induced using a $400\Gauss$ field~\cite{Taichenachev_Magnetic_2006} resulting in the same clock Rabi frequencies for cooling and spectroscopy as for \ybf.
Nonetheless, we find that cooling proceeds slower and 20 cooling cycles are needed to reach similar temperatures [Fig.~\subfigref{3}{a}] with a motional ground state fraction of $\gsf{2}=0.915^{+5}_{-5}$ ($\nbar{x}=0.044^{+4}_{-2}$, $\nbar{y}=0.047^{+4}_{-5}$).
We attribute the observed difference in the cooling performance to the different scattering properties of the two isotopes.
While fermionic \ybf is essentially non-interacting, bosonic \yb has a rather large scattering length of $105~a_0$~\cite{Kitagawa_Two-color_2008}, where $a_0$ is the Bohr radius.
This leads to a significant redistribution of thermal energy between the vertical and horizontal modes, counteracting cooling along the lattice axes.
This is further confirmed by a reduction of the vertical temperature, as observed via time-of-flight measurements -- an effect that is absent for the fermionic isotope [Fig.~\subfigref{3}{b}].
We further observe non-thermal sideband spectra at intermediate temperatures in the case of \ybf, due to the absence of rethermalization~\cite{SM}.
Rethermalization, on the other hand, offers a novel route for cooling \yb: while only one direction is directly cooled via CSC, the remaining directions undergo indirect cooling through scattering.
While we were able to demonstrate this technique~\cite{SM}, this process is relatively slow and comparable to conventional evaporative cooling methods.

The efficiency of our chirped CSC is mainly limited by repumping via the $\tD{1}$ state, which induces heating due to non-magic trapping conditions and atom loss induced via leakage to the anti-trapped \tP{2} state [Fig.~\subfigref{1}{a}], which both can be mitigated in the future by realizing a coherent two-photon repumping scheme that couples \tP{0} and \tP{1}.
This avoids population in \tP{2} and spontaneous decay from \tP{1} to \sS{0} can be made near-magic in the $759\nm$ lattice.
We characterize recoil heating to $\simeq\!0.08$ motional quanta per repumping cycle and measure total atom loss of about $30\%$ for the coldest 2D sideband spectra presented above.
The latter is consistent with an independent characterization that also accounts for additional loss due to inelastic collisions~\cite{SM}.
We further observe technical heating induced by the lattices of $\lesssim0.1$ quanta$/$s.

To demonstrate cooling in 3D, we employ \ybf as it does not require large magnetic fields~\cite{Taichenachev_Magnetic_2006}.
However, we expect similar cooling performance for both isotopes, particularly since thermalization plays only a minor role in 3D due to low site occupations.
To implement 3D CSC, we add a $\simeq\!2800\Erec$ deep vertical lattice (Lamb-Dicke parameter $\eta_z\simeq\!0.31$) to the $300\Erec$ deep horizontal lattices and interleave vertical cooling sweeps. The horizontal chirped CSC remains unchanged and the carrier Rabi frequency for the vertical direction is $\Omega/(2\pi)=4.6\kHz$ for cooling and $\Omega'/(2\pi)=1.3\kHz$ for spectroscopy.
Specifically, each cycle now consists of four frequency sweeps: one for each horizontal axis and two for the vertical direction.
The latter contains 30 cooling pulses, each $150\us$ in duration, and the cooling frequency is swept from $-15\kHz$ to $-33\kHz$.

We record sideband spectra for each direction after 15 cooling cycles (Fig.~\ref{fig:4}).
In the horizontal directions [Fig.~\subfigref{4}{a}] we find that the shape of the sideband differs from the one in 2D (Fig.~\ref{fig:2}).
Accurately capturing this shape with the theoretical sideband function requires precise knowledge of the relative populations of the large-spacing layers of the vertical lattice, which we cannot measure independently.
Instead we employ an asymmetric Voigt profile, which captures the blue sideband shape well.
Precise quantitative analysis of the red sideband is further complicated by reduced signal-to-noise (cf.\ detection noise in gray) and a potentially non-thermal population distribution.
We extract the mean motional occupation from the sideband asymmetry and find $\nbar{x}\simeq0.17$ and $\nbar{y}\simeq0.12$.
These are higher than in 2D CSC which we attribute to a hotter initial ensemble [insets of Fig.~\subfigref{4}{a,b}] and the additional recoil heating from cooling along the vertical direction.

The vertical cooling also effectively removes motional excitations in the out-of-plane direction [Fig.~\subfigref{4}{b}].
After cooling, the sideband spectrum reveals an additional structure, which
we attribute to the individual vertical layers.
Due to tightly-focused beams, the vertical lattice consists of only five trapped layers and their trap frequencies are sufficiently distinct that multiple separate peaks can be observed within the blue sideband (indicated by arrows).
We note that this effect necessitates chirped CSC even in a 1D geometry.
While we find moderate success in modeling this sideband structure during 1D spectroscopy (Fig.~\ref{fig:vertical_cooling}), here we extract $\nbar{z}\simeq0.16$ using the same procedure described above for simplicity.
This yields a 3D ground-state fraction of $\gsf{3}\approx66\%$, which is remarkable considering the high initial temperature.
Atom losses during cooling are severe ($\simeq\!90\%$), but can be mitigated as discussed above or by preparing lower initial-state temperatures. 
Nonetheless, comparing different number of cooling cycles suggests continued cooling beyond 15 cycles would further increase the ground state fraction.
We anticipate that removing the vertical trap frequency inhomogeneity, either by populating only a single layer or utilizing a retro-reflected lattice, would result in 3D CSC performance equivalent to the performance in 2D.

We illustrate the effect of 3D cooling for quantum applications by comparing Rabi oscillations on the clock transition (after optical pumping) prior to cooling and after cooling [Fig.~\subfigref{4}{c}].
While contrast is rapidly lost due to thermal dephasing for the hot ensemble, the CSC-cooled atoms show long-lived Rabi oscillations with a fitted $1/e$ contrast decay time of $11.4(2)\ms$.
Presently this is limited by the spatial inhomogeneity of the Rabi frequency across the cloud, as confirmed by a spatial analysis, and can be mitigated by increasing the size of the clock beam.
For the lowest $\nbar{}$ of $0.013$ achieved here, thermal dephasing results in a calculated clock $\pi$-pulse infidelity of $8\times10^{-5}$.
Sensitivity to spatial inhomogeneity of the Rabi frequency is reduced in a 1D geometry; indeed we find even longer coherence times of up to $\simeq\!30\ms$ after rethermalization cooling of \yb, as described in Ref.~\cite{SM}.

We have demonstrated high-fidelity ground-state cooling using the optical clock transition, which is applicable to all AELA independent of their nuclear spin.
The obtained ground state fraction is on par with other methods~\cite{Li_3D_2012, Klusener_Long-lived_2024, Norcia_Iterative_2024} and we have outlined a direct path to further improve the demonstrated cooling efficiency.
We highlight that, once atom loss during cooling is mitigated, the entropy achieved by 3D CSC is sufficiently low that adiabatic deformation of the lattice potential into a harmonic trap would result in degenerate Bose gases~\cite{Olshanii_Producing_2002} offering a novel route for fast all-optical cooling without the need for evaporation~\cite{stellmer_laser_2013,hu_creation_2017,urvoy_direct_2019,solano_strongly_2019,xu_bose-einstein_2024}.
This constitutes an exciting new avenue for bosonic AELAs with vanishing hyperfine structure where conventional sub-recoil laser cooling methods fail.
Fast assembly and ground-state initialization is crucial for enhancing the robustness and data rate of neutral atoms for all state-of-the-art applications in quantum simulation of itinerant particles, quantum information and metrology~\cite{Campbell_Fermi-degenerate_2017,phelps_sub-second_2020,young_tweezer-programmable_2022}.
The technique of CSC is particularly promising for applications that rely on multiple isotopes including double-isotope arrays~\cite{Nakamura_Hybrid_2024} and atom interferometers~\cite{Abe_Matter-wave_2021,abdalla_terrestrial_2025}.

\begin{acknowledgments}
We thank Leonardo Bezzo and Etienne Staub for experimental assistance, and Christoph Hohmann from MCQST for generating the 3D rendering.
This project has received funding from the Deutsche Forschungsgemeinschaft (DFG, German Research Foundation) under Germany’s Excellence Strategy -- EXC-2111 -- 390814868 (MCQST), via Research Unit FOR5522 under project number 499180199 and via Research Unit FOR5688 under project number 521530974.
We further acknowledge funding from the European Research Council (ERC) under the European Union’s Horizon 2020 research and innovation program (grant agreement No.~803047), from the German Federal Ministry of Education and Research (BMBF) via the funding program quantum technologies – from basic research to market (contract number 13N15895 FermiQP) and from the Initiative Munich Quantum Valley from the State Ministry for Science and the Arts as part of the High-Tech Agenda Plus of the Bavarian State Government.
This work has further received funding under Horizon Europe programme HORIZON-CL4-2022-QUANTUM-02-SGA via the project 101113690 (PASQuanS2.1) and from the European Union's Horizon 2020 Research and Innovation Programme under Grant Agreement no.~731473 and 101017733 (DYNAMITE).
R.M.K.\ additionally acknowledges support from the Alexander von Humboldt Stiftung.
\end{acknowledgments}

%


\cleardoublepage
\section*{Supplemental Material}
\renewcommand{\thefigure}{S\arabic{figure}}
\renewcommand{\theHfigure}{S\arabic{figure}} 
\renewcommand{\theequation}{S.\arabic{equation}}
\renewcommand{\thesection}{S.\Roman{section}}
\renewcommand{\thetable}{S\arabic{table}}

\setcounter{figure}{0}
\setcounter{equation}{0}
\setcounter{section}{0}
\setcounter{table}{0}

\section{Experimental sequence}
\label{sec:exp_sequence}
The loading sequence is similar for both isotopes and is the same for all data sets presented in this work.
It starts by loading a 3D MOT on the $\sS{0}\leftrightarrow\tP{1}$ transition for $500\ms$, which typically results in a total atom number of $\simeq2\times10^6$ for \yb and $\simeq1\times10^6$ for \ybf.
At the beginning of the MOT compression stage, in which the magnetic field gradient is ramped up and the detunings of the cooling beams are adjusted correspondingly to reach a denser and colder atomic cloud ($\simeq\!10\uK$), we suddenly turn on the 3D clock-magic lattice to $\simeq60\Erec$ in the horizontal direction and $\simeq6000\Erec$ in the vertical direction for \yb, while we use much deeper horizontal lattices of $\simeq600\Erec$ for \ybf, where $\Erec=h^2/(2m\lambda^2)$ is the recoil energy.
The highest loading efficiencies are obtained if all lattice beams are polarized in the horizontal $xy$-plane for \ybf.
For \yb the polarization of the horizontal lattice beams is changed to vertical.

In order to avoid significant population of the atoms in the wings of the 3D lattice, we perform a spilling sequence.
To this end, we turn off the horizontal lattices while simultaneously increasing the vertical lattice to $\simeq6700\Erec$ within $10\ms$.
After a short equilibration time of $10\ms$ we ramp the lattices to their final values used for cooling in 2D and 3D lattice configurations.
In the following we provide more details about the respective experimental sequences used for cooling, sideband spectroscopy and imaging.

\textit{2D-sideband cooling for \ybf}:
After isolating the crossing region, the horizontal lattices are ramped to $\simeq\!300\Erec$ each, while turning off the vertical confinement in $50\ms$.
To define $z$ as the quantization axis, we apply a magnetic field of $1.4\Gauss$ within $5\ms$.
Note that there is also a small $\simeq\!130\,$mG magnetic field in the $xy$ plane.
Details of the subsequent cooling pulse sequence are described in Sec.~\ref{sec:cooling_seq}.
The single-tone cooling sequence presented in Fig.~\subfigref{2}{a} of the main text is described in Sec.~\ref{sec:single_tone_cooling}.

\textit{2D-sideband cooling for \yb}:
As with \ybf, we gradually turn off the vertical confinement over $50\ms$ while ramping up the horizontal confinement to $\simeq\!300\Erec$ per lattice beam.
In contrast to fermions, bosons require a strong magnetic field to induce the clock transition~\cite{Taichenachev_Magnetic_2006}.
Hence, a magnetic field of $\simeq\!400\Gauss$ is applied along $z$ within $100\ms$.
The cooling pulse sequence is identical to that for \ybf and is described in Sec.~\ref{sec:cooling_seq}.

\textit{3D-sideband cooling for \ybf}:
Compared to the sequence for 2D-sideband cooling, the vertical lattice is left on but reduced to $\simeq\!2800\Erec$ while the horizontal lattices are turned on to $\simeq\!300\Erec$ in $50\ms$.
To ensure we can drive the clock transition with all three beams needed for 3D cooling, we apply a magnetic field of $2\Gauss$ with equal components in the $xy$ plane and along the $z$ axis.
We verified that this rotation of the quantization axis does not negatively affect the previously optimized 2D sideband cooling performance.

\textit{Sideband spectroscopy:}
Thermometry via sideband spectroscopy is typically performed in the same lattice configuration as the clock sideband cooling (CSC) in order to avoid heating induced by changing the lattice depth or geometry.
The carrier Rabi frequency $\Omega'=2\pi\times1.12\kHz$ of the spectroscopy pulse is low enough to avoid significant power broadening of the spectral features.
The pulse duration of $20\ms$ is chosen such that the sidebands are saturated and we can assume the spectrum as time-independent.
In this way, we ensure a sideband spectrum that can be reliably used for accurate thermometry.
We assume that the spectroscopy pulse does not affect the motional occupation of the atoms, which is justified given the long lifetime of the excited clock state $^3$P$_0$.
For wider transitions, the excited state can decay during probing which has to be carefully modeled~\cite{Norcia_Microscopic_2018}.
However, we note that the probe pulse duration can affect the inferred radial temperature~\cite{Blatt_Rabi_2009}.
For sideband spectroscopy along the $x$ and $y$ direction, the magnetic field is oriented along $z$ as for the 2D CSC.
Spectroscopy along the vertical axis is instead performed with a $1.4\Gauss$ magnetic field aligned parallel to the polarization of the vertical clock laser beam, to selectively drive $\pi$-transitions ($\Delta m = 0$).

\textit{Rabi oscillations:}
In order to drive full-contrast Rabi oscillations of \ybf in a 3D optical lattice, we spin-polarize the atoms via optical pumping to $\ket{F=1/2, m_F=1/2}$.
This takes place in the same lattice potential as 3D CSC, but the magnetic field is ramped up to $15\,\Gauss$ along $z$ within $100\,\ms$ to spectrally resolve the target optical pumping transition within the \tP{1} manifold.
Subsequently, the atoms are illuminated for $10\,\ms$ with one of the MOT beams to optically pump them to the desired state.
We then drive Rabi oscillations to $\ket{F'=1/2, m_{F'}=1/2}$ utilizing the clock beam along $y$, in the same lattice and magnetic field setting as the optical pumping step, with Rabi frequency $\Omega = 2\pi\times 3.7 \kHz$.

\textit{Imaging:}
Imaging is performed similar to our previous work~\cite{Hohn_Determining_2024}, where we leverage fluorescence imaging on the $\sS{0}\leftrightarrow\sP{1}$ transition enhanced by molasses cooling under a magic-angle condition on the $\sS{0}\leftrightarrow\tP{1}$ transition for fast, high signal-to-noise detection.
For \yb, the parameters are identical to those in Ref.~\cite{Hohn_Determining_2024}.
For \ybf, the internal level structure modifies the magic condition of the molasses cooling on the ${\sS{0}\leftrightarrow\tP{1}}$ transition.
A magic angle for $\ket{F'=3/2,m_{F'}=\pm1/2}$ of $\simeq\!17^\circ$ has been identified in a $759\nm$ potential~\cite{Lis_Midcircuit_2023}, but cannot be satisfied here for all three lattice beams simultaneously given their polarizations.
Instead, we identify a near-magic condition for the $\ket{F'=3/2,m_{F'}=\pm3/2}$ states at $90^\circ$, similar to the near-magic configuration found at $783.8\nm$ in Ref.~\cite{Norcia_Iterative_2024}.

The imaging is rendered state-selective by first removing remaining \sS{0} atoms with a pulse of resonant light on the $\sS{0}\leftrightarrow\sP{1}$ transition.
The pulse duration is $10\ms$ ($20\ms$) in a 2D (3D) lattice, which is sufficient to remove all ground state atoms.
This is confirmed by recording a second image, which yields photon counts consistent with a sequence where no atoms are loaded into the 3D MOT to begin with.
The observed background count is slightly higher for the 3D cooling sequences, which we ascribe to the vertical clock beam illuminating the camera and whose counts are not perfectly removed; this background is calibrated independently and subtracted from the 3D data in Fig.~\ref{fig:4} of the main text.
The shot-to-shot standard deviation in this background, assessed over 33 repetitions, is shown as a gray area in the same figure indicating our detection limit.
After removal of the \sS{0} population, the shelved \tP{0} atoms are repumped via the \tD{1} state with a $3\ms$ pulse, and subsequently imaged as above.

Time-of-flight and atom number measurements described in Secs.~\ref{sec:loss_heating} and \ref{sec:TOF} employ absorption imaging via $\sP{1}$.

\section{Experimental setup}
\label{sec:exp_setup}
\textit{Lattice laser beams:} The 3D clock-magic lattice is generated with three independent laser beams.
The 2D horizontal confinement is generated by two orthogonal retro-reflected 1D lattices, which are focused onto the atoms with waists of approximately $40\um$.
The vertical confinement is provided by two interfering laser beams from the side, which are generated using a K{\"o}sters prism and are horizontally polarized for maximal interference contrast.
The beams have a half-opening angle of $\theta = 9.75^\circ$ between them, generating vertical layers spaced by $\simeq\!2.2\um$.
The waists of the vertical lattice beams are colocated at the point where they interfere, and are elliptical, $26.6\um$ by $8.7\um$ with the long axis in the $xy$ plane, as characterized optically before integrating it in the main experimental setup~\cite{Hohn_State-dependent_2024}.

\textit{Clock beams:} The three clock beams are generated from the same laser source.
The beam is split into three and sent to the atoms via independent optical fibers without any active fiber noise cancellation.
Both of the horizontal clock beams are roughly co-propagating with the corresponding lattice axes, while the vertical beam propagates along $z$.
The clock beam propagating along $x$ ($y$) has a design waist of $132\um$ ($165 \um$), while the vertical beam's design waist is $173\um$.
The alignment onto the atomic cloud is done via spatial analysis of Rabi oscillations, in such a way that the average Rabi coupling is maximized and the center of the atomic cloud has the highest local Rabi frequency.
Both horizontal clock beams are linearly polarized along $z$, while the vertical clock beam is linearly polarized in the $xy$-plane.

\textit{Repumper beam:}
The repumper beam is nearly co-propagating with the lattice (and clock) beam along $x$, and has a design waist of $\simeq\!370\um$ and a power of $\simeq6\mW$.

\section{Data acquisition and analysis}
\label{sec:analysis}
Sideband spectroscopy is performed following the sequence described in Sec.~\ref{sec:exp_sequence}.
To extract an excitation fraction for each recorded spectrum, we measure the total atom number using an identical sequence but without applying the spectroscopy pulse and the removal of \sS{0} atoms.
(The data in Figs~\subfigref{2}{a} and \subfigref{sideband_2d_boson_single_tone}{a} is normalized by the total atom number prior to cooling.)
For each spectrum we randomize the clock-laser detunings and repeat the measurement multiple times as stated in the figure captions; error bars indicate the standard error of the mean over these repetitions.

In this Section we provide additional details about the fitting routine applied to the measured sideband spectra.
For spectroscopy in a 2D lattice, we perform an unweighted least-squares routine using Eq.~\eqref{eq:spectrum} including terms up to $\Delta n = 2$, i.e., the second sideband, and where each sideband is spatially averaged over multiple lattice sites, as given by Eq.~\eqref{eq:sideband2d_cheap}.
With the independently determined beam waist $w_0$ and cloud size $\sigma$, the remaining fit parameters are the lattice depth $V_x=V_y\equiv V$, longitudinal temperature $T_x$ or $T_y$, temperature in the weakly-confined direction $T_z$, and amplitudes $A_{0,1}$.
The amplitude of the second sideband is set as $A_2 = A_1/2$, and the carrier linewidth $\Gamma_0$ is fixed, the sideband linewidths are then $\Gamma_1 = \eta \Gamma_0$ and $\Gamma_2 = \eta^2 \Gamma_0 / \sqrt{2}$ where $\eta$ is the Lamb-Dicke factor.
From the optimal fit parameters, specifically the lattice depth and longitudinal temperature, we then calculate a mean motional occupation $\nbar{}$ and ground state fraction $\gsf{1}$ by including only states up to $n=N_\text{max}$, where $N_\text{max}$ is the number of trapped bands for a given lattice depth.
The 2D ground state fraction $\gsf{2}$ follows from multiplying the two $\gsf{1}$'s.
We note that all metrics are calculated and reported for the central lattice site: due to the lattice depth and trap frequency inhomogeneity, other lattice sites will have slightly different motional excitation fractions.
Additionally, we verified that these metrics are insensitive to significant, $\gtrsim25\%$ changes in the fixed parameters $w_0$, $\sigma$, and $\Gamma_0$.

For spectroscopy in a 3D lattice, we employ a simpler functional form for fitting, as mentioned in the main text.
Specifically, we use a spectrum given by
\be\begin{split}
    P(\tilde\omega) = \frac{A_0}{1 + \tilde\omega^2/\Gamma^2} &+ A_{-1} \mathcal{V}(\tilde\omega - \omega_0; \epsilon, \Gamma, a)\\
    &+ A_{+1} \mathcal{V}(-\tilde\omega - \omega_0; \epsilon, \Gamma, a),
\end{split}\ee
where $\tilde\omega$ is the probe detuning from the free-space resonance, $\omega_0$ is a proxy for the trap frequency, and
\be
    \mathcal{V}(\omega; \epsilon, \Gamma, a) = \mathcal{V}\big(\omega; S(\omega, \epsilon, a), S(\omega, \Gamma, a)\big)
\ee
is the asymmetric Voigt profile, defined in terms of the symmetric Voigt profile $\mathcal{V}(\omega; \epsilon, \Gamma)$ with Gaussian standard deviation $\epsilon$ and Lorentzian half-width at half maximum $\Gamma$, and the logistic function $S(x, A, a) = 2A(1+e^{ax})^{-1}$.
All parameters are free fit parameters, and are optimized via an unweighted least-squares routine, like above.
For these fits, the 1D ground state fraction is extracted from $A_{\pm1}$ using Eq.~\eqref{eq:area_to_gsf} and $\nbar{}$ follows from inverting $\gsf{1}=(1+\nbar{})^{-1}$.

All stated confidence intervals follow from a bootstrap analysis.
Specifically, we resample (with replacement) the individual experimental runs comprising an entire averaged sideband spectrum.
This bootstrap sample is then fitted as described above and we repeat this procedure 100 times for each spectrum.
The confidence intervals for each metric are then set by the percentile range that corresponds to the central $68.2\%$ of the probability, i.e., the range between the 16th and 84th percentile.
For evaluating the bootstrap error of the 2D ground-state fraction $\gsf{2}$ we take all possible combinations between both of the 1D ground state fractions along $x$ and $y$ to assess the probability distribution.

\section{Chirped CSC sequence}
\label{sec:cooling_seq}
As described in the main text, we employ pulsed cooling on the $\sS{0}\leftrightarrow\tP{0}$ clock transition.
The basic building block for this CSC consists of a pulse on the $\fixedket{\sS{0},n}\leftrightarrow\fixedket{\tP{0},n-1}$ transition followed by a $250\us$ repumper pulse that removes entropy from the system and (ideally) returns atoms to the ground state while preserving the motional quantum number.
This pair of pulses is repeated multiple times.
During these repetitions, the frequency of the clock laser is linearly swept, such that each pulse samples a fraction of the full frequency sweep and addresses those lattice sites that are resonant with that frequency interval.

\begin{figure}[t!]
	\includegraphics[width=\columnwidth]{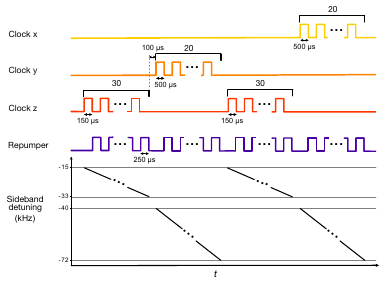}
	\caption{\label{fig:pulse_fig}
		\textbf{Clock sideband cooling sequence.} Illustration of one ``cooling cycle'' for 3D CSC consisting of cooling and repumping pulses combined with linear cooling frequency sweep.
        For 2D CSC the sequence is the same but the vertical cooling is omitted.}
\end{figure} 

The full cooling sequence for 3D CSC is illustrated in Fig.~\ref{fig:pulse_fig}.
The cooling sequence for 2D CSC is the same but omits all vertical cooling, using only the horizontal clock beams.
For those, each clock-repumper pulse pair is repeated 20 times, first for the clock-cooling beam along $y$, then repeated for the clock-cooling beam along $x$.
We observe no significant dependence of cooling performance on the time-ordering of the two beams.
Reducing the number of cooling pulses per frequency sweep (with fixed sweep range) results in worse cooling performance.
We also find that explicitly cooling both arms is necessary, unlike in tweezer arrays where one beam can suffice~\cite{Lis_Midcircuit_2023,Norcia_Iterative_2024}, since aberrations at the tweezers' focus typically result in a trap with two principal axes that are in general not aligned with the wave vector of the sideband cooling beam.
In lattices, a similar situation could be realized with a single clock laser beam propagating at a finite angle with respect to both lattices.
Note, however, that in this case the temperature cannot be characterized independently along each lattice axis.

The combined cooling of both horizontal directions constitutes one full cooling cycle.
We execute up to 15 (20) such cycles for \ybf (\yb).
Since the two horizontal lattices are equally deep, the clock pulse parameters are kept identical for the two beams.
Specifically, we employ a carrier Rabi frequency of $\Omega= 2\pi\times3.7\kHz$, limited for \yb by the maximum available laser power.
The cooling pulse duration of $500\us$ is chosen empirically to yield robust ground-state cooling.

The range of the frequency sweep is chosen to minimize the number of cycles required to reach the absolute ground state, which we experimentally find to be from $-40$ to $-72\kHz$.
This range is kept constant throughout the sequence.
Cooling performance appears identical for both sweep directions.
In this work we employ a downward sweep as shown in Fig.~\ref{fig:pulse_fig}.
Note that this is in contrast with direct sideband cooling in non-magic traps, where the chirp direction is governed by the sign of the relative light shift between the two states~\cite{Berto_Prospects_2021, Holzl_Motional_2023}.
However, as stated in the main text, for \ybf we observe non-thermal sidebands at intermediate numbers of cooling cycles, since atoms experiencing trap frequencies smaller than $40\kHz$ are not addressed and rethermalization is absent.
This can be mitigated by choosing a wider sweep range (experimentally verified with a sweep from $-20$ to $-72\kHz$), but this results in slower overall cooling and necessitates the application of more pulses.
This suggests improved performance with adaptive sweeps, where the chirp varies as a function of the number of cooling cycles.
Other variations, such as nonlinear frequency sweeps and non-square pulses, might further improve the cooling performance~\cite{Rasmusson_Optimized_2021}.

3D CSC is similar to 2D cooling with the key difference that in addition the vertical direction is cooled twice per cycle (Fig.~\ref{fig:pulse_fig}), because cooling in the vertical direction is inherently more challenging due to the relatively large layer-to-layer sideband frequency difference.
A single cooling period of the vertical direction consists of 30 clock pulses.
Their duration is $150\us$, the carrier Rabi frequency is $\Omega/(2\pi)=4.6\kHz$ ($\Omega'/(2\pi)=1.3\kHz$ for spectroscopy) and the sweep range extends from $-15\kHz$ to $-33\kHz$, which was empirically chosen to optimize the cooling performance for the available laser power.

\section{Ineffectiveness of single-tone cooling}
\label{sec:single_tone_cooling}
As demonstrated in the main text, single-tone cooling, i.e., sending the clock cooling pulses at a fixed detuning, is ineffective in higher dimensions due to the non-negligible inhomogeneity across the lattice sites/tubes.
To demonstrate this, the 2D CSC sequence described in Sec.~\ref{sec:cooling_seq} is modified by removing the frequency sweeps.
Additionally, we use a total of 40 cooling pulses in each direction, alternating between $x$ and $y$.
All other cooling parameters, i.e., power and duration of the clock and repumping pulses remain unchanged.

\begin{figure}[t!]
	\includegraphics{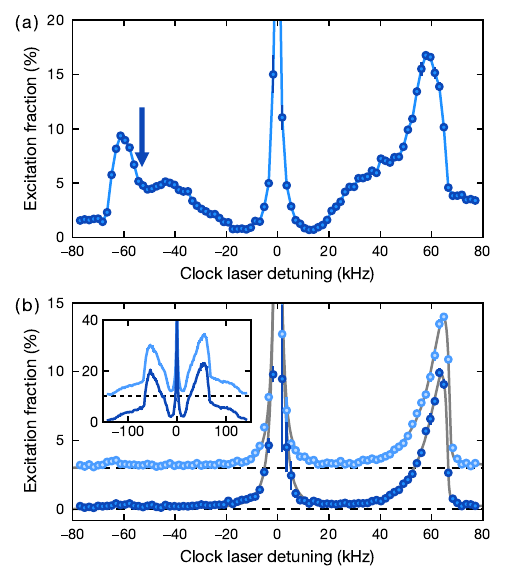}
    \caption{\label{fig:sideband_2d_boson_single_tone}
		\textbf{CSC of \yb in a 2D lattice}.
        \panel{a} Sideband spectrum probed by the clock beam along $y$ after single-tone cooling for \yb in the 2D lattice.
        The arrow indicates the chosen cooling frequency on the red sideband.
        The solid line connects the datapoints and is a guide to the eye.
        Error bars in both panels represent the standard error over three repetitions.
        \panel{b} Sideband spectrum after chirped CSC.
        Spectra along $x$ (light blue) and $y$ (dark blue) are taken after 20 cooling cycles, the former is vertically offset (dashed lines indicate zero) for better visibility.
        The gray solid lines are fits with the model discussed in the main text.
        Inset: spectra of initial state before cooling.}
\end{figure} 

We note that this effect is not unique to fermions in a 2D lattice.
We observe the same ineffectiveness for single-tone cooling of bosonic \yb in 2D, as shown in Fig.~\subfigref{sideband_2d_boson_single_tone}{a}.
However, the dip in the red sideband at the cooling frequency is less pronounced than for \ybf [Fig.~\subfigref{2}{a} in the main text], which is due to rethermalization between the bosons, see also Sec.~\ref{sec:rethermalization}.
The coldest spectra with 20 cycles of chirped CSC are shown in Fig.~\subfigref{sideband_2d_boson_single_tone}{b}.

For atoms in a 3D lattice, the inhomogeneity across the atomic ensemble is even more severe and we expect single-tone cooling to be even more ineffective.
Also, atoms in a 1D lattice formed with an interfering-beam geometry can require chirped CSC, as we demonstrate in Sec.~\ref{sec:vert_cooling}.
Finally, we note that this effect can even be observed in a retro-reflected 1D geometry, provided the lattice is sufficiently deep.
There we again find that single-tone CSC only addresses a narrow fraction of the sideband around the cooling frequency, this time not because of site-to-site inhomogeneity but rather due to cross-coupling with motional excited states in the radial direction.

\section{Atom loss and heating}
\label{sec:loss_heating}
This Section describes the characterization of loss and heating mechanisms that affect the CSC performance, as well as methods to mitigate these effects.

\subsection{Atom loss during cooling}
We monitor atom loss during CSC by measuring the total atom number versus the number of cooling cycles.
Regardless of the nature of CSC, i.e., 1D, 2D, or 3D, we observe similar behavior: initially atom loss is significant, but saturates as cooling progresses and practically vanishes once the motional ground state is reached.
This loss can be understood from several underlying mechanisms, all related to atoms in \tP{0}.
We note that the lifetime of atoms in \sS{0} is much longer than any experimental timescale, even for hot atoms.

\begin{figure}[t!]
	\includegraphics{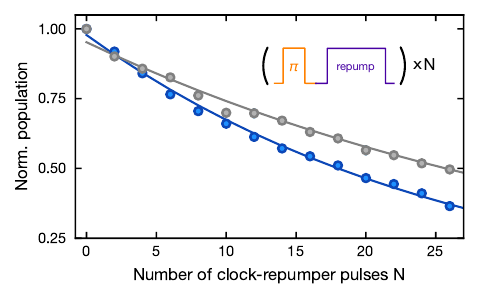}
	\caption{\label{fig:repumper_loss}
		\textbf{Repumper-induced atom loss.}
        Atom number when repeatedly cycling \yb atoms on the clock transition.
        The pulse sequence consists of a clock $\pi$-pulse followed by a repumper pulse, as illustrated in the inset.
        Data in gray (blue) is taken in a 3D (2D) lattice geometry, solid lines are fits with an exponential decay.
        Data points are the average over three repetitions, their standard error is smaller than the points.}
\end{figure}

We now experimentally quantify these losses.
During cooling, the number of atoms excited to \tP{0} by any given pulse is unknown.
Instead, we modify the sequence and excite a known number of atoms (i.e., the entire population) to \tP{0} using a clock $\pi$-pulse after 2D CSC, achieving near-unity transfer.
Subsequently, these atoms are immediately repumped using a $3\ms$ pulse, and the remaining atom number is detected.
The rest of the sequence, including the lattice configurations and chirped CSC, is identical to that described above.
To enhance the losses we repeat this process a variable number of times.
We note that the observed atom number does not precisely follow an exponential decay, especially when the clock and repumper pulses are repeated many times.
This is due to a reduced excitation efficiency during the $\pi$-pulse, as a result of heating as discussed below.
Hence, a fraction of these atoms remain in the ground state and do not participate in the repumping process, so the inferred loss rate is artificially lowered.
To avoid this issue we fit only the first 26 cycles, where a single exponential decay captures the data well (Fig.~\ref{fig:repumper_loss}).

The dominant loss mechanism is due to leakage via spontaneous emission from \tD{1} during the repumping process.
Despite its favorable branching ratios, a small fraction of atoms decays from \tD{1} to \tP{2} during the spontaneous emission, which is anti-trapped in the $759\nm$ lattice and directly lost.
The theoretical probability for this loss compared to successful repumping to \sS{0} is $\Gamma_{\tD{1}\to\tP{2}}(\Gamma_{\tD{1}\to\tP{2}} + \Gamma_{\tD{1}\to\tP{1}})^{-1}\simeq 2.5\%$.
Indeed, when performing the experimental characterization in a 3D lattice (vertical confinement of $\simeq\!1100\Erec$ ramped up within $25\ms$ after 2D CSC) we observe 2.47(6)\% atom loss as shown in gray in Fig.~\ref{fig:repumper_loss}.
Removing the additional vertical lattice, we observe a slightly larger loss fraction of $3.66(5)\%$ (blue data in Fig.~\ref{fig:repumper_loss}).
We ascribe these additional losses to inelastic collisions between \tP{0} atoms.
Their effect becomes more severe at higher density and lower temperature; we estimate approximately 1 atom per tube of the 2D lattice and significantly less than 1 atom per site in the 3D lattice.
Based on the observed losses in 2D, and the initial $\nbar{}$, we estimate an atom loss of $30\%$ for ground-state cooling, which is consistent with our observations.

A final mechanism that can affect atoms in \tP{0} is Raman scattering induced by the lattice photons, which causes state transfer to \tP{1} and \tP{2}.
The former radiatively decays to \sS{0} and atoms remain trapped, whereas the latter leads to atom loss, as described above.
A detailed quantitative description of Raman scattering for ytterbium can be found in Refs.~\cite{Siegel_Excited-Band_2024, Hohn_Determining_2024}.
We find that for typical lattice depths employed here, the $1/e$ lifetime of atoms in \tP{0} is much longer than the cooling duration.

\subsection{Repumper recoil heating}
\label{sec:repump_recoil}
In every repumping process, there are multiple uncontrolled photon recoils that can result in heating via momentum diffusion.
To measure the resulting heating rate, we use a similar sequence as above for repumper-induced atom loss, but instead of measuring the total atom number we perform sideband thermometry.
Specifically, we excite the atoms using the clock beam along $y$ and perform clock sideband spectroscopy with the clock beam along $x$, i.e., the same direction as the repumper beam.
We measure sideband spectra for different numbers of repeating clock-repumper pulse pairs and extract the average motional occupation by fitting them as described in Sec.~\ref{sec:analysis}.
We find a heating rate of 0.077 motional quanta per repumping cycle.

We note that both repumper-induced atom loss and recoil heating are not a fundamental limitation to CSC.
By employing a two-photon process, i.e., combining the $1388\nm$ laser with a $1539\nm$ laser driving the $\tP{1}\leftrightarrow\tD{1}$ transition, population (and thus loss) of the \tP{2} state can be avoided, and recoil heating can be suppressed by a judiciously chosen orientation of the two repumper beams.
Alternatively, one can employ modified cooling schemes where each atom only experiences one repumping pulse to reach the ground state~\cite{Li_3D_2012}.
Finally, a triple-magic lattice configuration for \sS{0}, \tP{1} and \tP{0} can mitigate recoil heating due to the remaining uncontrolled decay from \tP{1} to \sS{0}~\cite{Hohn_Determining_2024}.

\subsection{Lattice-induced heating}
Phase and intensity noise of the lattice light can cause the atoms in the optical trap to heat up, i.e., increase the motional occupation number.
High-resolution sideband spectroscopy after cooling to the motional ground state is ideally suited to characterize such heating; we quantify it by holding the atoms for a variable amount of time after CSC.
The observed heating rate depends on the precise operating conditions, but is found to be less than 0.1 motional quanta per second.

\section{Additional spectroscopy after 3D clock sideband cooling}
In addition to the sideband spectra presented in Fig.~\ref{fig:4} of the main text, we perform further experiments to ascertain the occupation of the 3D motional ground state.
All experiments described here follow the same 3D cooling sequence as in the main text, but differ in the steps after cooling.
These are aimed at either improving the signal-to-noise ratio by reducing the depth of the 3D spectroscopy lattice, or allow us to fit the theoretically expected sideband shapes for lower-dimensional lattices to the measured sideband spectra by modifying the dimensionality of the lattice for spectroscopy.
Both these approaches confirm a large 3D motional ground state fraction.

\begin{figure*}[t!]
	\includegraphics{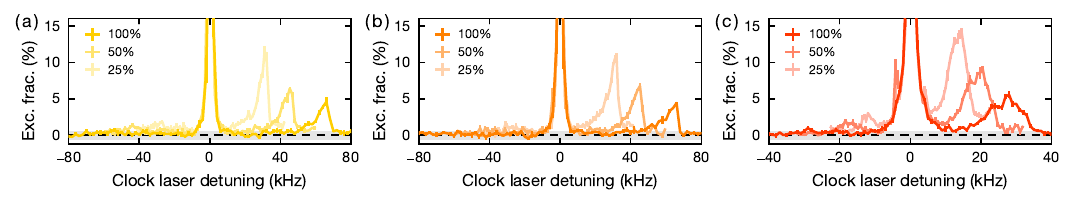}
    \caption{\label{fig:sideband_alternative_3d}
	\textbf{Sideband spectra after 3D CSC versus spectroscopy lattice depth}.
    Spectroscopy along $x$, $y$ and $z$ is illustrated in \panel{a}, \panel{b} and \panel{c}, respectively, and shows the expected reduction of sideband frequency and increased signal strength.
    The legends indicate the fraction of the spectroscopy lattice depth compared to the cooling potential.
    Error bars indicate the standard error over five repetitions.
    The gray shaded area indicates our detection limit.}
\end{figure*} 

For the first set of experiments, we adiabatically lower the trap depth of all three lattices utilizing a linear ramp over $50\ms$ to a final setpoint where spectroscopy is performed.
The shallower 3D lattice has a reduced sideband frequency, which increases the Lamb-Dicke factor and thus the sideband Rabi frequency.
As a result, the area under each sideband increases, improving the signal-to-noise ratio at the cost of reduced resolvability of higher-order sidebands.
We compare spectroscopy in the original, full-depth lattice to spectra obtained at $50\%$ and $25\%$ of the original lattice depth.
The results are presented in Fig.~\ref{fig:sideband_alternative_3d}, where we indeed observe an increased prominence of the sidebands for all three directions, approximately consistent with the expected $V^{-1/4}$ scaling, where $V$ is the depth of the lattice along the spectroscopy axis.
The sideband is compressed but retains its shape, as expected.
A quantitative analysis using fits with the same asymmetric lineshape as in Fig.~\ref{fig:4} yields ground state fractions for each direction that are consistent with those reported in the main text.

The second set of experiments involves changing the trapping geometry after cooling, wherein the sideband spectra are more easily modeled.
In particular, we adiabatically ramp off either the two horizontal lattice beams yielding a 1D vertical lattice, or vice versa, using an exponential ramp with time constant $250\us$ and total duration $1.75\ms$ (thus reaching $1/e^7$ of the original depth before any remaining light is quenched off).
For the lattice direction(s) where confinement remains constant, we expect preservation of the motional state while for the other direction(s) heating may occur, e.g., due to slightly non-adiabatic ramps.
We thus expect that the sideband spectra provide accurate information about the temperature and motional state occupation in the original lattice along the corresponding probe direction.
Further, the spectra in the reduced geometry are amenable to fitting with the simpler 1D and 2D functional forms, where our models are applicable.
Indeed, we find good correspondence between data and fit, especially for the two horizontal spectra.
From the fit we extract $\nbar{x}=0.092$, $\nbar{y}=0.086$, and $\nbar{z}=0.43$, for a 3D ground state fraction of $\gsf{3}=59\%$, which is slightly lower than the values reported in the main text.

\section{Time-of-flight thermometry}
\label{sec:TOF}
As a cross-check to our sideband thermometry experiments, we also perform time-of-flight (TOF) thermometry.
We note that interpretation of TOF thermometry is more complicated when releasing the atoms from a lattice potential~\cite{McKay_Lattice_2009}, especially for fermions.
However, it additionally allows us to characterize the temperature along the weakly confined directions, for which sideband spectroscopy is unreliable if thermal equilibrium is not reached, as discussed in Sec.~\ref{sec:thermometry}.
We thus utilize it to gain qualitative insight into the underlying cooling and heating processes.

\begin{figure}[ht!]
	\includegraphics{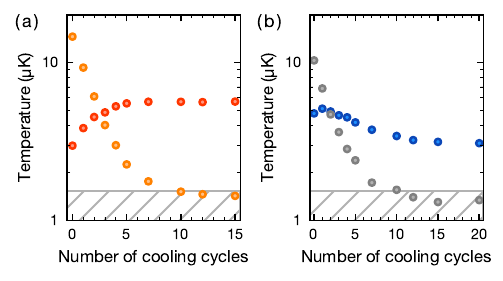}
	\caption{\label{fig:quench_TOF}
		\textbf{Time-of-flight thermometry after 2D cooling}.
        \panel{a,b} Temperature versus number of cooling cycles after sudden release from the lattice, evaluated from fits of the time-of-flight evolution of the cloud size.
        The fit uncertainty is smaller than the data points.
        The hatched area indicates the temperature of the calculated average zero-point-motion in the 2D lattice configuration.
        \panel{a} Horizontal (orange) and vertical (red) temperature of 2D-cooled \ybf atoms.
        \panel{b} Horizontal (gray) and vertical (blue) temperature of 2D-cooled \yb atoms.}
\end{figure} 

A first set of experiments rapidly quenches off the lattice confinement after 2D CSC; the resulting expansion is fitted with the functional form $\sigma(t) = \sqrt{\sigma(0)^2 + k_B T t^2/m}$, where $\sigma(t)$ is the $e^{-1/2}$ radius of the density profile at time $t$ after release, $k_B$ is the Boltzmann constant, and $T$ is the temperature.
For atoms in the motional ground state, the observed temperature is limited by the zero-point motion in the lattice.
Indeed, for both \ybf and \yb the cooled, horizontal direction appears to saturate close to the calculated limit, shown by the hatched area in Fig.~\ref{fig:quench_TOF}.
The calculated limit here is based on a minimal model that neglects interatomic interactions and the reduction in longitudinal confinement resulting from finite radial temperature, but accounts for the spatial distribution of atoms across lattice sites.

As discussed in the main text, the behavior along the uncooled, weakly confined vertical direction differs between the isotopes.
For \ybf, the temperature along this direction rises due to recoil heating caused by the repumping process (Sec.~\ref{sec:repump_recoil}), which becomes progressively more rare as the 2D horizontal motional ground state is reached.
In contrast, for \yb, the vertical temperature increases only initially due to this effect, but then gradually reduces as collisions redistribute thermal energy to the actively-cooled horizontal directions.

In order to adiabatically remove the zero-point energy that limits the above experiments, we perform further TOF thermometry where the lattice is exponentially ramped down, rather than quenched.
Ideally this reveals any motional excitation above the zero-point energy in the lattice.
The duration and the characteristic time of the exponential ramp-down are chosen in such a way that the process is adiabatic: for the 2D lattice (after 2D CSC), the duration of the ramp is $2\ms$ with a time constant $0.5\ms$, whereas after 3D CSC these parameters are $1.75\ms$ and $0.25\ms$, respectively.
Due to gravity, the atoms fall out of the lattice before the end of these ramps; we track the evolution of cloud size after the entire ramp is completed.
There, we observe a linear growth of the cloud size, which we fit with the asymptotic form of the functional form above, i.e., $\sigma(t) = \sqrt{k_B T/m}(t-t_0)$ where $t=0$ is the end of the ramp and $t_0<0$ captures the effective time of release.
After completing 2D CSC, i.e., 15 (20) cycles for \ybf (\yb), the horizontal direction shows virtually no expansion, with a fitted temperature of $\simeq\!11\nK$ ($\simeq\!3\nK$) that is rather limited by systematics such as finite imaging resolution and finite field of view limiting the achievable TOF duration.
The vertical temperature observed in these experiments qualitatively matches that discussed above, i.e., slowly decreasing with cooling for \yb, and saturating at a slightly higher temperature for \ybf.
After 3D cooling, the cloud appears to fall as a rigid body, with no expansion to be observed in any direction, even after $10\ms$ TOF.
This supports a high-fidelity ground state occupation consistent with thermometry based on the sideband spectra presented in the main text in stark contrast to rapid expansion without any cooling.

\section{Rethermalization cooling}
\label{sec:rethermalization}
In the main text, we mention the possibility of exploiting the scattering properties of $\yb$ to passively cool the directions perpendicular to the CSC direction.
This technique further relies on a rather high density of atoms on each lattice site.
In 2D, the number of atoms per tube is not sufficient to facilitate many scattering events between atoms.
For this reason, we demonstrate this cooling method in a 1D magic lattice, where the number of atoms per layer is enough to observe efficient rethermalization.
The longitudinal direction is cooled using CSC, while the radial directions are cooled via rethermalization with the longitudinal direction.
We extract these two temperatures from fitted sideband spectra with the functional dependence described by Eqs.~\eqref{eq:spectrum}, \eqref{eq:sidebandKernel} and \eqref{eq:shape1d} of the main text.

Rethermalization is slow, but gets progressively more effective in deeper lattices.
At the same time, single-tone cooling becomes more ineffective in deeper trapping potentials (even in 1D) as discussed in Sec.~\ref{sec:single_tone_cooling}.
To address this we interleave cooling periods at the usual lattice depth ($\simeq\!300 \Erec$) with rethermalization periods at the maximum depth we can generate ($\simeq 2000 \Erec$).

For CSC we use 25 pulses, all at a fixed detuning from the carrier of $-64\kHz$.
In principle, this single-tone cooling has the issues described in Sec.~\ref{sec:single_tone_cooling}, but in addition to the relatively shallow lattice depth for cooling, this is mitigated by the fact that we repeat this cooling multiple times.
After cooling, the strong magnetic field that induces the transition is turned off for technical reasons.
At the same time, the lattice trap depth is linearly increased to $\simeq 2000 \Erec$ over $8.5\ms$, and held there for $70\ms$ to allow rethermalization to take place.
Then, in anticipation of the next cooling cycle the lattice is ramped back down and the strong magnetic field is again ramped up within $25\ms$.
This rethermalization cycle can be repeated, and sideband spectra are recorded each time immediately after the last set of cooling pulses.

\begin{figure}[t!]
    \centering
    \includegraphics{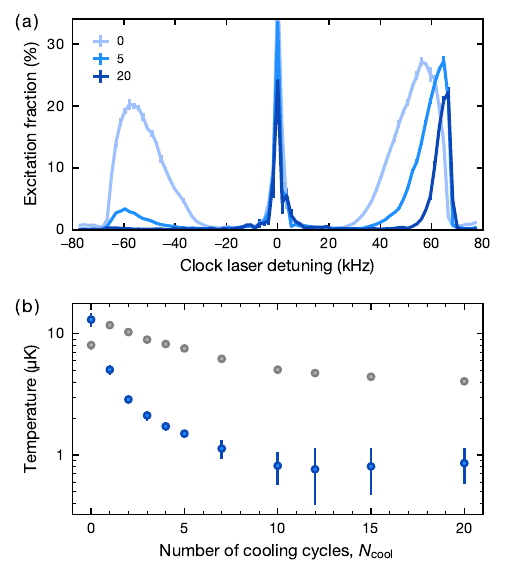}
    \caption{\label{fig:rethermalization}
    \textbf{Rethermalization cooling of \yb in a 1D lattice.}
    \panel{a} Sideband spectra for different numbers of rethermalization cooling cycles, as indicated in the legend.
    Both longitudinal cooling (vanishing red sideband) and radial cooling (narrowing blue sideband) are evident.
    Error bars indicate the standard error over three repetitions.
    \panel{b} Sideband spectra are fitted to extract longitudinal (blue) and radial (gray) temperatures versus the number of cooling cycles.
    Error bars correspond to the fit uncertainty.
    Rethermalization continues to passively cool the radial direction even when the longitudinal temperature has already saturated.}
\end{figure}

Figure~\subfigref{rethermalization}{a} shows three sideband spectra that illustrate the rethermalization effect.
In addition to suppression of the red sideband by direct CSC, the blue sideband gets narrower with more rethermalization cycles, which is an indication of the radial direction getting colder.
After fitting, we extract the longitudinal and radial temperatures versus number of rethermalization cycles, shown in Fig.~\subfigref{rethermalization}{b}.
Specifically, we count the number of cooling periods $N_\text{cool}$.
For instance, $N_\text{cool}=10$ means that CSC was applied ten times with nine periods of rethermalization in the deep lattice in between.
Thus, $N_\text{cool}=0$ corresponds to the directly loaded atomic cloud (no cooling), while $N_\text{cool}=1$ has longitudinal cooling but no rethermalization.
There the radial temperature rises due to repumper heating during longitudinal cooling, cf.\ Sec.~\ref{sec:loss_heating}.
For subsequent cycles, rethermalization redistributes energy from the radial to the longitudinal direction, the latter is then directly recooled via CSC.
We note that quantitative interpretation of radial temperatures from fitting sideband spectra is non-trivial: the probing during spectroscopy potentially affects the inferred temperature, as discussed in Refs.~\cite{Blatt_Rabi_2009, Wang_Effective_2022}.
Nonetheless, the radial direction is evidently cooled.
This rethermalization cooling scheme can be further optimized by, e.g., dynamically adjusting parameters such as the duration of the rethermalization period, sideband cooling frequency, etc.

\section{Cooling in a shallow-angle interference lattice}
\label{sec:vert_cooling}
The vertical lattice employed in our work is formed by interfering two focused beams at a shallow angle.
This forms a lattice with a small number of layers that further each have different trap frequencies, see also Sec.~\ref{sec:sideband_shallow-angle}.
This inhomogeneity is more severe the smaller the beams' waists.
In our setup only five layers are trapped.
As noted in the main text, this necessitates chirped CSC even in a 1D lattice of this kind, which we demonstrate here for \ybf.

To load atoms into the 1D vertical lattice, we follow the same procedure as described in Sec.~\ref{sec:exp_sequence}, where after preparation in a deep vertical lattice, its depth is reduced to the cooling depth of $V_z\simeq2800\Erec$ yielding a sideband frequency of $\simeq40\kHz$.
Additionally, to improve the initial population distribution we briefly lower the lattice depth to $\simeq\!100\Erec$ before returning it to the cooling depth.
This spills atoms in highly excited states out of the trap, resulting in $\simeq\!15.6\%$ of the population.
Cooling is then performed as described in Sec.~\ref{sec:cooling_seq}, with the same sweep to address all layers.

\begin{figure}[t!]
	\includegraphics{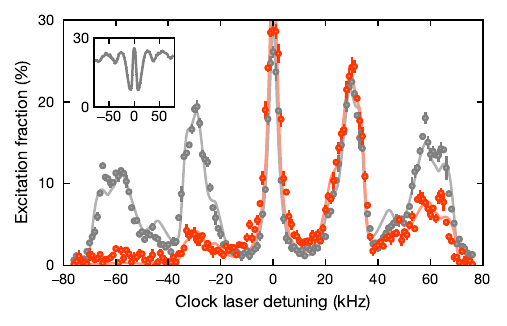}
	\caption{\label{fig:vertical_cooling}
		\textbf{Cooling of \ybf in a shallow-angle interference 1D lattice.}
        Sideband spectra prior to cooling (gray) and after 20 cycles of chirped CSC (red).
        The corresponding solid lines are fits that account for layer-to-layer trap frequency inhomogeneity, see text for details.
        Error bars indicate the standard error over five repetitions.
        Inset: spectrum without cooling and spilling.}
\end{figure}

Spectroscopy reveals the same sideband structure as discussed in the main text, which becomes particularly striking in the second sideband (Fig.~\ref{fig:vertical_cooling}).
The spectrum after 20 cycles of chirped CSC displays significant reduction of motional excitations compared to the spectrum prior to cooling (inset in Fig.~\ref{fig:vertical_cooling}).
Both spectra are fitted with the sideband theory developed in Sec.~\ref{sec:sideband_shallow-angle}, specifically using the sideband kernel given by Eq.~\eqref{eq:kernel_1dvertical}.
Here, we input the beam geometry parameters as found from an optical characterization, i.e., those listed in Sec.~\ref{sec:exp_setup}.
Additionally, we fix $\phi=144^\circ$ and set the layer populations approximately proportional to the layer's depth.
The remaining parameters (overall lattice depth, longitudinal and radial temperature, and sideband amplitudes) are then found from a least-squares optimization procedure.
Between various spectra we find that no change in, e.g., layer population fractions is needed to faithfully reproduce the observations, including the observed structure in the first and second sideband shapes arising from the layer-to-layer inhomogeneity.
This further indicates good stability of the relative phase $\phi$ between the two interfering laser beams.
From the fit we extract $\nbar{z}\simeq0.17$ after cooling, noting that this number characterizes the motional occupation in the central layer; the population-averaged result is $\nbar{z}\simeq0.22$.
This clearly illustrates the additional challenges faced from cooling in this lattice geometry: the inhomogeneities that arise necessitate chirped CSC which cools more slowly.

Lastly, we remark on the use of the additional spilling step.
If this step is omitted, we observe a hot ensemble with barely resolvable sidebands (inset of Fig.~\ref{fig:vertical_cooling}).
We find that with such a challenging initial-state distribution, chirped CSC works but requires a much larger number of cycles (and hence substantial atom loss, cf.\ Sec.~\ref{sec:loss_heating}) to achieve low temperatures and significant motional ground state fractions.
This can be understood as a consequence of the system exceeding the Lamb-Dicke limit: Atoms in highly excited motional states $n$ violate the Lamb-Dicke condition $\eta^2(2n+1)\ll1$, despite the Lamb-Dicke factor $\eta$ being small.
Consequently, these atoms cannot be efficiently cooled by sideband cooling.
This effect might also affect cooling performance for the 3D CSC presented in the main text, where spectra prior to cooling are similarly hot.

\section{Theory of sideband spectra}
This Section provides the analytical theory describing sideband spectra.
We first describe a general framework before specifying the sideband shapes in various cases.

\subsection{General framework}\label{sec:sideband-framework}
We focus on neutral atoms confined by optical potentials probed on an ultranarrow transition.
Compared to, e.g., harmonically trapped ions where the sidebands are narrowly peaked~\cite{Leibfried_Quantum_2003}, anharmonicities in the optical potential render the sideband a smooth distribution.
For 1D optical lattices the shape of the sidebands has been discussed in Refs~\cite{Blatt_Rabi_2009, McDonald_Thermometry_2015}.
Here we summarize the approach of Ref.~\cite{Blatt_Rabi_2009}, formalize some of the approximations used therein and extend the model to higher dimensions.

The atoms are interrogated with a probe beam with wavevector $\mathbf{k}$ and frequency $\omega=c\abs{\mathbf{k}}$, detuned from the free-space resonance $\omega_{ge}$ by $\tilde{\omega} = \omega - \omega_{ge}$.
If the probe pulse is sufficiently long in duration, we can assume a quasi time-independent treatment.
Then, for a transition at frequency $\omega_{\Delta\mathbf{n}}=\omega_{ge}+\tilde\omega_{\Delta\mathbf{n}}$, the excitation probability is given by the incoherent lineshape
\be\label{eq:lineshape}
	p_e(\tilde\omega_{\Delta\mathbf{n}}, \tilde\omega)=L_\Gamma(\tilde\omega)* \delta\left(\tilde\omega - \tilde\omega_{\Delta \mathbf{n}}\right),
\ee
where $L_\Gamma(\tilde\omega)= \frac{1}{2} \frac{1}{1+4\tilde\omega^2/\Gamma^2}$ and $\delta(\tilde\omega)$ is the Dirac delta-function.
This assume a maximum excitation fraction of 0.5 and $\Gamma$ is the linewidth which arises due to, e.g., finite lifetime of the excited state or power broadening of the probe light.
The spectrum will consist of a summation of this lineshape over all possible transitions.
Eigenfunctions of the 3D trapping potential are indexed by the three quantum numbers $\mathbf{n}^\nu = (n^\nu_x,n^\nu_y,n^\nu_z)$, with $\nu=\{g,e\}$, so that the entire spectrum is described by
\be
	P(\tilde\omega) = \sum\limits_{\mathbf{n}^g,\mathbf{n}^e=0}^N f_T(\mathbf{n}^g) \cdot p_e(\tilde\omega_{\Delta\mathbf{n}}, \tilde\omega),
\ee
where $\mathbf{n}^\nu$ denotes the set of quantum numbers for the ground and excited state and $\Delta\mathbf{n}=\mathbf{n}^e-\mathbf{n}^g$ denotes the corresponding transition.
The summation runs over all quantum numbers, $N$ is a generic notation for the upper bound of the summation and is equal to the number of trapped states (which can be different for each summation), and $f_T(\mathbf{n}^g)$ describes the initial distribution of atoms.
However, not all transitions will be excited, depending on the (relative) orientation of the probe beam and trapping potential.
This is determined primarily by the Lamb-Dicke parameters,
\be
	\eta_i = \sqrt{\frac{\hbar(\mathbf{k}\cdot\hat{\mathbf{e}}_i)^2}{2m\omega_i}},
\ee
where $m$ is the atomic mass and $\omega_i$, $i=\{x,y,z\}$, the trapping frequency.
The Rabi frequency for a transition depends on the Lamb-Dicke parameters via
\be\label{eq:Rabi}
	\frac{\Omega(\mathbf{n}^g,\mathbf{n}^e)}{\Omega_0} = \prod\limits_{i=x,y,z}e^{-\eta_i^2/2}\sqrt{\frac{n_i^<!}{n_i^>!}}\eta_i^{\abs{n_i^e-n_i^g}}\mathcal{L}^{(\abs{n_i^e-n_i^g})}_{n_i^<}(\eta_i^2),
\ee
where $\Omega_0$ is the free-space Rabi frequency, $n_i^{>(<)}$ is the maximum (minimum) of $n_i^g$ and $n_i^e$ and $\mathcal{L}_n^{(\alpha)}(x)$ is the generalized Laguerre polynomial~\cite{Wineland_Laser_1979, Leibfried_Quantum_2003}.
As a result, if $\eta_i=0$ then only transitions between equal quantum numbers $n^g_i = n^e_i$ are allowed.
Thus, if we assume that the probe beam is copropagating with one of the principal directions of the trapping potential, the Lamb-Dicke parameters in the perpendicular directions vanish.
For the following we denote the principal axis as $z$.
Hence, only transitions with $\Delta\mathbf{n} =(0,0,\Delta n)$ have non-zero Rabi frequencies and we label the corresponding transition frequencies as $\omega_{\Delta n}$.
Grouping the transitions by $\Delta n$ simplifies the expression for the excited state fraction as follows:
\be\begin{split}\label{eq:sideband_grouping}
	P(\tilde\omega) &= \sum\limits_{\Delta n}\sum\limits_{\substack{\mathbf{n}^g=(0,0,n_z^\text{min})}}^N f_T(\mathbf{n}^g)\cdot p_e(\tilde\omega_{\Delta n}, \tilde\omega) \\
    &\equiv \sum\limits_{\Delta n} P_{\Delta n}(\tilde\omega),
\end{split}\ee
where $n_z^\text{min}=\max(-\Delta n,0)$ to ensure that $n_z^g+\Delta n\geq0$.
This grouping is primarily mathematical, but can be understood as the confinement of the optical potential resulting in a separation of the spectrum into distinct sidebands indexed by $\Delta n$ and given by $P_{\Delta n}(\tilde\omega)$.
The lower bound $n_z^\text{min}$ thus correctly reflects that the first $n$ motional states are dark states when driving the $n$-th red sideband.

As we will show below, the shapes of the different sidebands can be directly related to each other.
We thus focus on the first blue sideband, given by $\Delta n=+1$, with shape
\be\label{eq:sideband_exact}
	P_{+1}(\tilde\omega) = \sum\limits_{\mathbf{n}=0}^N f_T(\mathbf{n}) \cdot p_e(\tilde\omega_{+1}, \tilde\omega).
\ee
Note that $\tilde\omega_{\Delta n}$ depends on the initial motional state, and we suppress the superscript $g$ for notational convenience.
While in general each transition is characterized by a different lineshape that depends on the quantum numbers via Eq.~\eqref{eq:Rabi}, this only weakly affects the shape of the spectrum and we assume that there is a single linewidth $\Gamma_{+1}$ for all transitions within the blue sideband, allowing us to write
\be\label{eq:sideband_kernel}
	P_{+1}(\tilde\omega) = L_{\Gamma_{+1}}(\tilde\omega) * \underbrace{\sum\limits_{\mathbf{n}=0}^N f_T(\mathbf{n})\cdot \delta\left(\tilde\omega - \tilde\omega_{+1}\right)}_{\rho_{+1}(\tilde\omega)},
\ee
where from now on we will refer to $\rho_{+1}(\tilde\omega)$ as the sideband kernel.
Note that Ref.~\cite{Blatt_Rabi_2009} effectively only considers the sideband kernel, rather than $P_{+1}$ which also accounts for the finite linewidth of each individual transition.
Also note that the kernel is area-normalized, $\int\limits_{-\infty}^\infty \rho_{+1}(\tilde\omega) \text{d}\tilde\omega =1$, since $\sum\limits_{\mathbf{n}=0}^N f_T(\mathbf{n})=1$ by definition.

The expression above for the sideband kernel still involves a triple sum, which is numerically impractical and not revealing of any qualitative features or understanding.
To proceed, we utilize the fact that usually the sideband frequency $\tilde\omega_{+1}$ depends weakly on at least one quantum number.
For the purposes of illustration, we pick this to be $n_x$.
By weak dependence we mean that $\abs{\tilde\omega_{+1}(n_x+1) - \tilde\omega_{+1}(n_x)} \ll \Gamma_{+1}$, such that the summation over this number results in a smooth sideband shape $P_{+1}(\tilde\omega)$.
In that case, for the sideband kernel we can approximate the summation with its Riemann integral, resulting in
\be\begin{split}\label{eq:rho_integral}
	\rho_{+1}(\tilde\omega) &\approx  \sum\limits_{n_y,n_z=0}^N \int\limits_0^N \text{d}n_x f_T(\mathbf{n})\cdot \delta\left(\tilde\omega - \tilde\omega_{+1}\right)\\
    &=\sum\limits_{n_y,n_z=0}^N f_T(\tilde{n}_x)\left(\frac{\partial \tilde\omega_{+1}}{\partial n_x} \biggr|_{\tilde{n}_x}\right)^{-1} \Theta(\tilde{n}_x),
\end{split}\ee
where we used the compounded sifting property of the Dirac delta-function, and $\tilde{n}_x$ is the solution to $\tilde\omega = \tilde\omega_{+1}(\tilde{n}_x)$.
Here we assume there is only a single such solution, if there are multiple solutions these should be summed over.
The result is a simplified summation, and one that will elucidate the role of the various temperatures, as we will see below when using this framework to derive specific sideband shapes.
Finally we note that the probe direction should not be eliminated this way, because of its role in the relation between sidebands as we explain below.

\subsection{Relation between sidebands}
Before discussing sideband shapes for specific trapping geometries, we first clarify the relation between the different sidebands.
Assuming a magic trapping potential~\footnote{Or a two-photon Raman transition, where the two states tend to automatically have identical trapping potentials}, the ground and excited state experience the same optical potential $V(x,y,z)$.
We are interested in the eigenenergies of this potential as these determine how the transition frequency is modified compared to the free-space resonance at $\omega_{ge}$.
Typically, the potential is well-approximated by a harmonic expansion, at least near its minimum.
To fully capture the sideband spectrum we further need to consider anharmonicities that couple different directions of the trap.
Expanding the trap to quartic order and treating the quartic terms using a perturbation theory, the eigenenergies can be written generically as
\be\begin{split}\label{eq:Equartic}
	E(\mathbf{n}) = &\sum\limits_{i=x,y,z}\left[\hbar\omega_i\left(n_i+\frac{1}{2}\right) + \hbar\Omega_{ii}\left(n_i^2+n_i+\frac{1}{2}\right)\right]\\
	& + \sum\limits_{\langle i,j\rangle}\hbar\Omega_{ij}\left(n_i+\frac{1}{2}\right)\left(n_j+\frac{1}{2}\right),
\end{split}\ee
where the first line is the energy of each 1D harmonic oscillator with quartic anharmonicity, and the second line sums over the non-identical (unordered) pairs of coordinate directions with $\langle i,j\rangle = \{xy, xz, yz\}$.
Since this energy is a quadratic form, the transition frequency
\be
	\omega_{\Delta\mathbf{n}} = \omega_{ge} + \frac{1}{\hbar}\left[E(\mathbf{n}^e) - E(\mathbf{n}^g)\right],
\ee
is as well.
In particular, for spectroscopy along $z$, the sidebands of interest are indexed by $\Delta\mathbf{n}=(0,0,\Delta n)$ and we can derive a simple expression for all higher-order sideband frequencies as a function of the first blue sideband with $\Delta n=+1$:
\begin{equation}
\tilde{\omega}_{\Delta n} =\omega_{\Delta n}-\omega_{ge} = \Delta n\, \tilde{\omega}_{+1} + \Delta n(\Delta n-1)\Omega_{zz}
\end{equation}
and 
\begin{eqnarray}
\tilde{\omega}_{0}&=&\omega_{0} - \omega_{ge} = 0 \nonumber\\
\tilde{\omega}_{+1}&=&\omega_{+1}-\omega_{ge} \nonumber\\
&=&\omega_z + 2\Omega_{zz} (n_z+1) + \sum_{i=x,y} \Omega_{iz}\left(n_i+\frac{1}{2}\right).
\end{eqnarray}
This allows us to relate the different sideband shapes to that of the kernel of the first blue sideband, since with Eq.~\eqref{eq:sideband_grouping} we can write
\be\begin{split}
	P_{\Delta n}(\tilde\omega) &= L_{\Gamma_{\Delta n}}(\tilde\omega) * \sum\limits_{\substack{\mathbf{n}^g=(0,0,n_z^\text{min})}}^N f_T(\mathbf{n})\cdot  \delta\left(\tilde{\omega} - \tilde{\omega}_{\Delta n}\right) \\
    &=\frac{1}{\abs{\Delta n}}L_{\Gamma_{\Delta n}}(\tilde\omega) * \rho_{(n_z^\text{min})}\left(\frac{1}{\Delta n}\tilde{\omega}-(\Delta n-1)\Omega_{zz}\right),
\end{split}\ee
where we again used a single linewidth $\Gamma_{\Delta n}$ to describe all lineshapes constituting the sideband, and defined the generalized sideband kernel
\be
	\rho_{(n_z^\text{min})}(\tilde{\omega})\equiv \sum\limits_{\substack{\mathbf{n}^g=(0,0,n_z^\text{min})}}^N f_T(\mathbf{n})\cdot \delta (\tilde{\omega} - \tilde{\omega}_{+1}),
\ee
which explicitly accounts for the lower bound of the summation.
Note that $\rho_{+1}(\tilde{\omega}) \equiv \rho_{(0)}(\tilde{\omega})$.
These manipulations do not hold for $\Delta n=0$, i.e., the carrier, but there the spectral shape is much simpler thanks to the magic condition: $P_0(\tilde\omega) = L_{\Gamma_0}(\tilde\omega)$.
Combining these results, the total spectrum can be expressed as
\be\begin{split}
	P(\tilde\omega) &= \sum\limits_{\Delta n}P_{\Delta n}(\tilde\omega) \\
	&=L_{\Gamma_0}(\tilde{\omega}) + L_{\Gamma_{+1}}(\tilde\omega) * \left[\rho_{(0)}(\tilde{\omega}) + \rho_{(1)}(-\tilde{\omega}+2\Omega_{zz})\right]\\
	&\hspace{1em}+\frac{1}{2}L_{\Gamma_{+2}}(\tilde\omega) \\
    &\hspace{2em}* \left[\rho_{(0)}\left(\frac{1}{2}\tilde{\omega}-\Omega_{zz}\right) + \rho_{(2)}\left(-\frac{1}{2}\tilde{\omega}+3\Omega_{zz}\right)\right] + \cdots
\end{split}\ee
where the terms are ordered by $\Delta n$, and we used that the Rabi frequencies for any red transition are essentially equal to their blue counterparts, so that $\Gamma_{\Delta n} = \Gamma_{-\Delta n}$.

Finally, we revisit the quasi time-independent treatment assumed throughout.
In writing the incoherent lineshape Eq.~\eqref{eq:lineshape}, we assumed saturation at 50\% excitation probability.
This is realized in the long-time limit, which might not be achieved for realistic probe pulse durations.
Empirically, the excitation probability at a given detuning typically grows asymptotically towards this value over time, as various nearby sideband transitions are off-resonantly excited and incoherently add (the carrier is instead described by decaying Rabi oscillations).
The timescale for the growth is related to the sideband Rabi rate, which in the Lamb-Dicke regime depends primarily on $\abs{\Delta n}$.
As before we assume this is uniform over the sideband so that it can be captured with a rescaling factor $A_{\abs{\Delta n}}$.
This then yields
\be\begin{split}\label{eq:spectrum_SM}
    &P(\tilde\omega) = L_{\Gamma_0}(\tilde{\omega}) + \sum\limits_{\Delta n = 1} A_{\Delta n} L_{\Gamma_{\Delta n}}(\tilde\omega) \\
    &* \left(\rho_{(0)}\left[\frac{\tilde{\omega}}{\Delta n}-(\Delta n-1)\Omega_{zz}\right] + \rho_{(\Delta n)}\left[\frac{-\tilde{\omega}}{\Delta n}+(\Delta n+1)\Omega_{zz}\right]\right),
\end{split}\ee
for which the first few terms are given in Eq.~\eqref{eq:spectrum} of the main text where we additionally inserted the typical $\Omega_{zz}=-\frac{1}{2}\rec$ for standard retro-reflected 1D lattices, with $\rec$ the recoil frequency.

\subsection{Thermometry}\label{sec:thermometry}
We briefly remark on the usual utility of sideband spectroscopy for thermometry purposes.
Results in this context are usually stated without qualifiers, which we clarify here.
Assuming what we have done above, primarily that the incoherent lineshape generated for each transition has identical width for all transitions composing a single sideband order, one finds
\be\label{eq:area_to_gsf}
	\frac{\mathcal{A}_{-1}}{\mathcal{A}_{+1}} = \frac{\sum\limits_{\substack{\mathbf{n}=(0,0,1)}}^N f_T(\mathbf{n})}{\sum\limits_{\mathbf{n}=(0,0,0)}^N f_T(\mathbf{n})} = 1 - \sum\limits_{n_x,n_y}^N f_T(n_z=0) \equiv 1 - F_0^z,
\ee
where $\mathcal{A}_{\Delta n}$ is the area under the sideband indexed by $\Delta n$ and $F_n^i$ is the population fraction in the $n$-th motional state along direction $i$.
Likewise, for the second sideband
\be
	\frac{\mathcal{A}_{-2}}{\mathcal{A}_{+2}} = 1 - F_0^z - F_1^z = \frac{\mathcal{A}_{-1}}{\mathcal{A}_{+1}} - F_1^z,
\ee
which contains the population fraction in the first motionally excited state.
Generally, the population distribution can be arbitrarily shaped, but in thermal equilibrium we can assume a Boltzmann distribution,	$f_T(\mathbf{n}) = Z^{-1}\prod\limits_{i=x,y,z}\exp(-E_i(n_i)/(k_BT_i))$, with $Z$ the partition function.
Evaluating this requires separating out the eigenenergy $E(\mathbf{n})$ into separate components, which can be done by e.g.\ ignoring the contributions to Eq.~\eqref{eq:Equartic} from the quartic order distortions of the potential.
Further assuming $N\rightarrow\infty$ we can then express the area ratios as
\be
	\frac{\mathcal{A}_{-m}}{\mathcal{A}_{+m}} = \exp(-m\beta_z) = \left(1 + \frac{1}{\bar{n}_z}\right)^{-m},
\ee
where $\beta_i = \frac{\hbar \omega_i}{k_BT_i}$ and $\bar{n}_i$ is the mean motional occupation number for direction $i$.
As such, sideband spectroscopy allows straightforward determination of the temperature along the direction of probing.
Information about temperatures along other directions can also be inferred, however, to obtain accurate values knowledge about the trapping potential has to be included, as we will detail in the following.

\subsection{Sideband shapes}\label{sec:sideband-shapes}
We now derive explicit expressions for the sideband shapes in various typical trapping potentials.
For each case we specify the potential, perform a quartic expansion and specify the coefficients of Eq.~\eqref{eq:Equartic}, and evaluate the sideband kernel $\rho_{+1}(\tilde{\omega})$ using the approach in Eq.~\eqref{eq:rho_integral}.
Where feasible we also analytically perform the Lorentzian convolution.
Throughout, we will assume Boltzmann weighting as above, define $\zeta_i=\exp(-\beta_i)$, assume static atoms (i.e., no tunneling), and assume $N\rightarrow\infty$ for the index we eliminate.

\subsubsection{Retro-reflected 1D lattice}
We assume a Gaussian laser beam with wavelength $\lambda$ propagating along $z$ that is perfectly retro-reflected onto itself.
The resulting trapping potential is
\be\label{eq:lattice1d}
	V(x,y,z) = - V_z \cos^2(kz)\text{e}^{-2(x^2+y^2)/w_0^2},
\ee
where $V_z$ is the lattice depth, $k=2\pi/\lambda$, $w_0$ is the beam waist, and we assume the typical scenario of the Rayleigh range $z_R=\pi w_0^2/\lambda$ being large compared to other length scales such as the extent of the atomic cloud, so that it can be ignored.
The minima of this potential are at $x=0$, $y=0$, and $z=l\lambda/2$ for $l\in\mathbb{Z}$.
Since the potential is translationally invariant along $z$, all these minima have identical shape and it suffices to focus on the minimum at $z=0$.
There, we can expand the potential to quartic order and obtain
\be
	\frac{V(x,y,z)}{V_z}\simeq -1 + k^2z^2 + \frac{2}{w_0^2}r^2-\frac{1}{3}k^4z^4-\frac{2k^2}{w_0^2}r^2z^2-\frac{2}{w_0^4}r^4,
\ee
where $r=\sqrt{x^2+y^2}$.
Diagonalizing the harmonic oscillator and performing perturbation theory on the quartic terms, we find the coefficients for the eigenenergies defined in Eq.~\eqref{eq:Equartic} as
\be\begin{split}\label{eq:energy_param_1d}
	&\Omega_{zz} = -\frac{1}{2}\rec,\\
	&\Omega_{xx}=\Omega_{yy}=-3\rec\omega_r^2/(4\omega_z^2),\\
	&\Omega_{xz}=\Omega_{yz}=-\rec\omega_r/\omega_z,\\
	&\Omega_{xy}=-\rec\omega_r^2/\omega_z^2,\\
    	&\omega_z=\sqrt{2k^2V_z/m}=2\rec \sqrt{\mathcal{V}_z},\\
	&\omega_x = \omega_y = \sqrt{4V_z/(mw_0^2)}=\omega_z\sqrt{2}/(kw_0)\equiv\omega_r,
\end{split}\ee
where $\hbar\rec=\Erec=\hbar^2k^2/(2m)$, and $\mathcal{V}_z = V_z/(\hbar\rec)$.
Thanks to the radial symmetry, in this geometry $x$ and $y$ are treated on equal footing.
We can thus define $n_r=n_x+n_y$ so that the sideband kernel from Eq.~\eqref{eq:sideband_kernel} becomes
\be
	\rho_{+1}(\tilde\omega) = \sum\limits_{n_r,n_z=0}^N (n_r+1) f_T(n_r,n_z)\cdot \delta\left(\tilde\omega - \tilde\omega_{+1}(n_r,n_z)\right),
\ee
where we also assumed that $T_x=T_y\equiv T_r$.
The sideband frequency relation is then given by
\be
	\tilde\omega_{+1}(n_r,n_z) = \omega_z - \rec(n_z+1) - \rec\frac{\omega_r}{\omega_z}(n_r+1).
\ee
Since $w_0\gg\lambda$ we have $\omega_r\ll\omega_z$, hence the sideband frequency weakly depends on $n_r$ and we will eliminate that according to the procedure outlined in Sec.~\ref{sec:sideband-framework}.
For that we solve $\tilde\omega = \tilde\omega_{+1}$ to find
\be\begin{split}
	\tilde{n}_r(\tilde\omega) &= \frac{\omega_z}{\rec\omega_r}\left(\omega_z - \tilde\omega - \rec(n_z+1)\right) - 1 \\
    &\equiv \frac{\omega_z}{\rec\omega_r}s_z(\tilde\omega;n_z)-1,
\end{split}\ee
where we defined the function $s_z(\tilde\omega;n_z)$ for convenience.
Hence, the sideband kernel is
\be\label{eq:kernel_radial}
	\rho_{+1}(\tilde\omega)\propto\sum\limits_{n_z=0}^N\left(\tilde{n}_r(\tilde\omega)+1\right)f_T(\tilde{n}_r(\tilde\omega),n_z)\,\Theta(\tilde{n}_r(\tilde\omega)).
\ee
Inserting the Boltzmann factors we obtain
\begin{align}\label{eq:kernel1d}
	&\rho_{+1}(\tilde\omega)\approx\frac{C}{Z_z}\sum\limits_{n_z=0}^N \zeta_z^{n_z}f_{1\text{D}}\left(s_z(\tilde\omega; n_z)\right),\\
    \label{eq:kernel1d_fun}
    &f_{1\text{D}}(s) = s e^{-\alpha s}\Theta(s),
\end{align}
with $Z_z=\frac{1-\zeta_z^{N+1}}{1-\zeta_z}$ is the partition function, $\alpha = \beta_r\omega_z/(\rec\omega_r)$ and where we ignored a small contribution $-\rec\omega_r/\omega_z$ inside the step-function.
The prefactor $C$ can be found by ensuring this result remains area-normalized, which yields $C = \alpha^2$.
The expression for the generalized kernel $\rho_{(n_z^\text{min})}(\tilde\omega)$ follows trivially by directly modifying the lower limit of the summation.
We point out that this expression matches the main result of Ref.~\cite{Blatt_Rabi_2009}, and additionally note that by writing out $\alpha$ in full, i.e., $\alpha = \hbar\omega_z/(\rec k_BT_r)$, the sideband kernel does not depend on $\omega_r$ at all, and hence not on $w_0$ either.
The only parameters that enter are $\omega_z$, $\rec$, $T_z$ and $T_r$.
This is somewhat peculiar: the cross-coupling between the $z$- and $r$-directions occurs naturally in a way that it exactly counteracts the $\omega_r$ brought in by the Boltzmann factor for the radial quantum states.

The final functional form used for fitting, i.e., Eq.~\eqref{eq:spectrum_SM}, requires convolving the kernel with a Lorentzian $L_\Gamma$ which we will now specify.
The sideband kernel Eq.~\eqref{eq:kernel1d} contains the fundamental function $f_{1\text{D}}(s)$ evaluated at $s_z(\tilde\omega; n_z)$ and summed over $n_z$.
The function $s_z(\tilde\omega; n_z)$ is a linear transformation of $\tilde\omega$, specifically a reflection and displacement.
Since the Lorentzian itself is reflection-symmetric, both of these operations commute with the convolution and it suffices to calculate the convolution for the underlying function itself.
The result of that is
\be
	I_{1\text{D}}(y) = \int\limits_{-\infty}^\infty \text{d}x f_{1\text{D}}(x) L_\Gamma(y - x) = \frac{\Gamma}{4\alpha}\Im\{z e^{-z} \Gamma(0,-z) \},
\ee
with $\Im\{\cdot\}$ denoting the imaginary part, $z = \alpha(y+i\Gamma/2)$ and $\Gamma(a,z)$ is the upper incomplete gamma function~\cite{Abramowitz_Handbook_1972}, defined for $\abs{\arg{z}}<\pi$.
The blue sideband shape is thus
\be\label{eq:sideband1d}
	P_{+1}(\tilde\omega) = \frac{\alpha^2}{Z_z}\sum\limits_{n_z=0}^N \zeta_z^{n_z} I_{1\text{D}}\left(s_z(\tilde\omega; n_z)\right),
\ee
where the remaining summation has to be performed numerically.
We note that we need not assume a Boltzmann distribution for the $z$-direction; indeed this final result holds for general population distributions $f_T(n_z)$ with the appropriate substitution of $\zeta_z^{n_z}$ and $Z_z$.

\begin{figure}[t!]
	\includegraphics{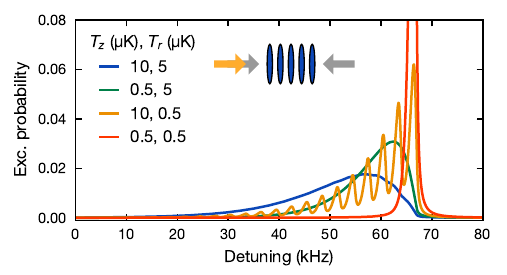}
	\caption{\label{fig:sideband_1d}
		\textbf{Sideband shape in 1D}.
		Illustration of the sideband shape, Eq.~\eqref{eq:sideband1d}, for different temperatures.
        For all curves we take $\omega_z = 2\pi\times70\kHz$, $\rec = 2\pi\times3\kHz$ and $\Gamma = 2\pi\times0.5\kHz$.
        The red curve is cropped for visibility reasons but consists of a single, sharp peak.
        Inset: Schematic of the lattice (gray arrows) and probe (yellow arrow) geometry.}
\end{figure}

Figure~\ref{fig:sideband_1d} illustrates the sideband shape for different parameter regimes.
Through $\alpha$, the radial temperature $T_r$ primarily determines the width of the sideband kernel.
Sidebands for different initial $n_z$ have nearly identical shape, and the longitudinal temperature $T_z$ determines how many `copies' play a role.
When $T_r$ is sufficiently low, the distinct longitudinal sidebands are resolvable, as shown in the orange curve in Fig.~\ref{fig:sideband_1d} and observed in Ref.~\cite{Zhang_Subrecoil_2022}.

Before considering other lattice geometries, we discuss retro-reflected 1D lattices generated by an elliptical beam.
In this case, the potential defined in Eq,~\eqref{eq:lattice1d} is modified and the Gaussian envelope is given by $\exp(-2x^2w_x^{-2} - 2y^2w_y^{-2})$.
This results in the following modifications compared to Eq.~\eqref{eq:energy_param_1d} of the parameters specifying the energy landscape:
\be\begin{split}
	&\omega_x = \omega_z \sqrt{2} / (k w_x),\\
	&\omega_y = \omega_z \sqrt{2} / (k w_y),\\
	&\Omega_{xz} = -\rec\omega_x/\omega_z,\\
	&\Omega_{yz} = -\rec\omega_y/\omega_z.
\end{split}\ee
We omit the changes in the expressions for $\Omega_{xx}$ and $\Omega_{yy}$ as they are irrelevant in this calculation.
The blue sideband frequency is now
\be\begin{split}
	\tilde\omega_{+1}(n_x,n_y,n_z) = \,&\omega_z - \rec(n_z+1) \\
    &- \frac{\rec}{2\omega_z}\left(\omega_x(2n_x+1) + \omega_y(2n_y+1)\right).
\end{split}\ee
The next step, of substituting $n_r = n_x+n_y$, no longer works, because the sideband frequency does not have this radial degeneracy.
However, we point out that if we again assume that $T_x=T_y\equiv T_r$, then the sideband kernel solely depends on the quantity $\omega_xn_x+\omega_yn_y\equiv\nu$, since the Boltzmann factor in that case is $\exp(-\hbar\nu/(k_BT_r))$.
For the isotropic 1D lattice considered above, introducing $n_r$ reduced the double summation over $n_x$ and $n_y$ into a single summation over $n_r$, which was then approximated by an integral.
We point out that equivalently, we could have instead first approximated the double summation with a double integral and then perform a change of variables for the integral to arrive at the same result.
This latter strategy still works here, as long as $\omega_x$ and $\omega_y$ are not too dissimilar and both result in sufficiently weak dependence of $\tilde\omega_{+1}$ on $n_x$ and $n_y$.
Following this strategy, we find
\be\begin{split}
	\rho_{+1}(\tilde\omega)\propto\sum\limits_{n_z=0}^N&\left(\tilde\nu(\tilde\omega) + \frac{1}{2}(\omega_x+\omega_y)\right)\\
    &\times f_T(n_z)\exp\left(-\frac{\hbar\tilde\nu(\tilde\omega)}{k_BT_r}\right)\Theta(\tilde\nu(\tilde\omega)),
\end{split}\ee
where $\tilde\nu(\tilde\omega)$ denotes the solution to $\tilde\omega_{+1}=\tilde\omega$ for $\nu$.
Specifically, that is
\be\begin{split}
	\tilde\nu(\tilde\omega) + \frac{1}{2}(\omega_x+\omega_y) &= \frac{\omega_z}{\rec}\left(\omega_z - \tilde\omega - \rec(n_z+1)\right) \\
    &= \frac{\omega_z}{\rec}s_z(\tilde\omega; n_z),
\end{split}\ee
where we identified the same expression $s_z(\tilde\omega; n_z)$.
Substituting this result for $\tilde\nu(\tilde\omega)$ yields precisely the same sideband kernel (and hence sideband shape) as that given in Eq.~\eqref{eq:kernel1d}.

\subsubsection{Shallow-angle 1D lattice}\label{sec:sideband_shallow-angle}
In the retro-reflected 1D lattice, the lattice spacing is fixed by the wavelength.
These can be decoupled by interfering two running-wave beams at an angle~\cite{Ville_Loading_2017}, such as employed for the vertical lattice in this work.
The resulting lattice structure also modifies the sideband spectrum, which we will now discuss.

\begin{figure}[t!]
	\includegraphics{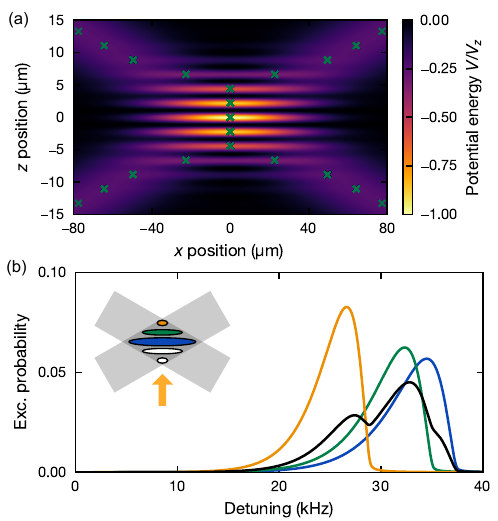}
	\caption{\label{fig:sideband_vertical}
		\textbf{Sideband shape for a shallow-angle interference lattice}.
		\panel{a} Illustration of the lattice potential given by Eq.~\eqref{eq:shallow-angle_potential}, for the vertical lattice used in this work, i.e.\ with parameters $w_H=26.6\um$, $w_V=8.7\um$, $\theta=9.75^\circ$, $\phi=0$, and $\lambda=759\nm$.
        Local minima of the potential are indicated by green crosses.
        \panel{b} The sideband shape depends on where inside the lattice the atoms reside: the colored curves correspond to different layers as indicated by the schematic inset.
        The black line is the average over an ensemble that occupies all 5 layers with approximately Gaussian population distribution, leaving the layers barely resolvable.
        The parameters for the lattice potential are identical to \panel{a}, with $\mathcal{V}_z=3000$ ($\simeq\!290\uK$), $T_z=15\uK$, $T_r=75\uK$, and probe Rabi frequency $\Omega = 2\pi\times0.5\kHz$ for the carrier (the sideband Rabi frequency is different for each layer).
        The arrow in the inset indicates the probe beam direction.
        }
\end{figure}

We consider two identical beams, both propagating in the $xz$-plane.
One has an angle to the $x$-axis of $+\theta$ while the other has $-\theta$.
Inspired by the experimental implementation, we consider elliptical rather than circular Gaussian beams, with waist in the $y$-direction of $w_H$ and waist in the $xz$-plane of $w_V$.
The resulting optical potential is given by
\be\begin{split}\label{eq:shallow-angle_potential}
	&\frac{V(x,y,z)}{V_z} = -\frac{1}{2}\exp\left(-\frac{2y^2}{w_H^2}-\frac{2x^2\sin^2(\theta)}{w_V^2}-\frac{2z^2\cos^2(\theta)}{w_V^2}\right)\\
    &\hspace{4em}\times\left[\cos(2kz\sin(\theta)-\phi) + \cosh\left(\frac{2xz\sin(2\theta)}{w_V^2}\right)\right],
\end{split}\ee
where $\phi$ is the relative phase between the two beams.
Here, the exponential describes an overall envelope, the first term inside the rectangular brackets is the lattice potential and the second term describes the dipole traps, which further determine the shape of the trap far away from the interference region.
We ignore the evolution of beam sizes due to the finite Rayleigh range for simplicity.
The minima of this potential are not immediately obvious, but can be numerically found and are depicted in Fig.~\subfigref{sideband_vertical}{a}.
We discern two types: those at $x=0$ and those away from $x=0$.
The former can be understood directly from the cosine term in the potential function: it creates trap minima at $z = a(l+\phi/(2\pi))$ where $a=\lambda/(2\sin(\theta))$ is the lattice spacing and $l\in\mathbb{Z}$.
However, not all integers $l$ result in local minima of the potential, as is clear from Fig.~\subfigref{sideband_vertical}{a}.
This can be understood by evaluating the Hessian of the potential at $x=0$, $y=0$.
When any of its eigenvalues (which directly relate to the three trap frequencies) changes sign, the extrema of the potential change from stable minima to unstable saddlepoints.
Setting the determinant of the Hessian equal to zero results in a transcendental equation that has to be solved for $z$.
It can be shown that the magnitude of the solution has an upper bound of $z_\text{max} = w_V / (\cos(\theta)\sqrt{2})$.
For $\abs{z}>z_\text{max}$ all local minima are thus of the second type, that is, away from $x=0$.
In what follows, we will assume these minima are not populated by atoms and hence do not need to be taken into account for the sideband spectrum.
The remaining minima will be referred to as `layers', and the maximum number of trapped layers is
\be
	N_l = \left\lceil\frac{2z_\text{max}}{a}\right\rceil = \left\lceil2\sqrt{2}\tan(\theta)\frac{w_V}{\lambda}\right\rceil.
\ee
Each of these layers has a different depth and local potential shape, and will need to be analyzed separately.
The result will be a sideband spectrum that accounts for this layer-to-layer inhomogeneity.
This is thus clearly distinct from the retro-reflected 1D lattice, where all minima were identical thanks to translational invariance.

We start by assuming $\phi=0$ and analyzing the central layer at $z=0$.
The shape of the potential for this layer is not so different from that of the retro-reflected 1D lattice, and after expanding to quartic order we find
\be\begin{split}
	&\omega_0 = 2\rec\sin(\theta)\sqrt{\mathcal{V}_z},\\
	&\omega_z=\omega_0\sqrt{1+\frac{2}{k^2w_V^2\tan^2(\theta)}},\\
	&\omega_x = \omega_0 \sqrt{2}/(kw_V), \\
	&\omega_y = \omega_0 \sqrt{2}/(kw_H\sin(\theta)),\\
	&\Omega_{zz} = -\frac{\rec\omega_0^2}{2\omega_z^2}\left(\sin^2(\theta) + \frac{6\cos^2(\theta)}{k^2w_V^2} + \frac{6\cos^4(\theta)}{\sin^2(\theta)k^4w_V^4}\right),\\
	&\Omega_{xz}=-\frac{\rec\omega_x}{\omega_z}\left(\sin^2(\theta)+\frac{4\cos^2(\theta)}{k^2w_V^2}\right),\\
	&\Omega_{yz}=-\frac{\rec\omega_z\omega_y}{\omega_0^2}\sin^2(\theta),
\end{split}\ee
where we defined $\omega_0$ for convenience and omitted irrelevant parameters.
Since $1/(kw_{V,H})\ll1$, going forward we will ignore all subleading orders in it, so that e.g.\ $\omega_z=\omega_0$ and $\Omega_{zz}=-\rec\sin^2(\theta)/2$, though we emphasize this is purely for notational convenience and does not impact the derivation.
The resulting sideband frequency is then
\be\begin{split}
	\tilde\omega_{+1}(n_x,&\,n_y,n_z) = \omega_z - \rec\sin^2(\theta)(n_z+1)\\
    &- \frac{\rec\sin^2(\theta)}{2\omega_z}\left[\omega_y(2n_y+1)+\omega_x(2n_x+1)\right],
\end{split}\ee
which is identical to that of the anisotropic retro-reflected lattice, but where $\rec$ is effectively renormalized by $\sin^2(\theta)$.
The derivation of the sideband kernel and shape then follows identically, yielding again Eqs.~\eqref{eq:kernel1d} and \eqref{eq:kernel1d_fun} for the kernel, but with $\alpha = \hbar\omega_z / (k_B T_r \rec\sin^2(\theta))$ and $s_z(\tilde\omega; n_z)$ replaced by 
\be
    s_\theta(\tilde\omega; n_z) = \omega_z - \tilde\omega - \rec\sin^2(\theta)(n_z+1).
\ee

Next, we analyze non-central layers.
These layers have shallower depths due to the $z$-evolution of the envelope function.
It can be shown that for a layer located at $z=z^*$, the trap frequencies can be related to that of the (fictitious, if $\phi\neq0$) $z=0$ layer via
\be\begin{split}
	&\omega_z(z^*) \approx \omega_z \exp\left(-(z^*\cos(\theta)/w_V)^2\right),\\
	&\omega_y(z^*) \approx \omega_y \exp\left(-(z^*\cos(\theta)/w_V)^2\right),\\
	&\omega_x(z^*) \approx \omega_x \exp\left(-(z^*\cos(\theta)/w_V)^2\right)\sqrt{1 - (z^*/z_\text{max})^2}.
\end{split}\ee
The exponential matches exactly the scaling expected from the reduced local trap depth, while the trap frequency along $x$ has a further modification that is related to the fact that the layers become unstable minima at $\abs{z}=z_\text{max}$.
The relevant quartic terms follow a similar scaling, though these are proportional to the local trap depth directly:
\be\begin{split}
    &\Omega_{zz}(z^*) \approx \Omega_{zz} \exp\left(-2(z^*\cos(\theta)/w_V)^2\right)\\
    &\Omega_{xz}(z^*) \approx \Omega_{xz} \exp\left(-2(z^*\cos(\theta)/w_V)^2\right)\\
    &\Omega_{yz}(z^*) \approx \Omega_{yz} \exp\left(-2(z^*\cos(\theta)/w_V)^2\right),
\end{split}\ee
where again $\Omega_{zz}$ etc., refers to that of the $z=0$ layer.
These scalings do not modify the derivation of the sideband kernel, and that of a non-central layer is given by the appropriate substitutions.
Explicitly, indexing the layers by integers $l$, we find the total kernel to be
\be\begin{split}\label{eq:kernel_1dvertical}
    \rho_{+1}(\tilde\omega) = &\sum\limits_l \frac{p_l\alpha_l^2}{Z_{z,l}} \sum\limits_{n_z=0}^N\zeta_{z,l}^{n_z}\\
    &\times f_{1\text{D}, l}\left(\omega_zr_l - \tilde\omega - \rec r_l^2\sin^2(\theta)(n_z+1)\right),
\end{split}\ee
where $p_l$ is the layer's population fraction, $\alpha_l = \hbar\omega_z/(k_BT_r\rec r_l\sin^2(\theta))$, $f_{1\text{D}, l}(s)$ is given by Eq.~\eqref{eq:kernel1d_fun} but with $\alpha$ replaced by $\alpha_l$, and $r_l = \exp(-\lambda^2(l+\frac{\phi}{2\pi})^2/(4\tan^2(\theta)w_V^2))$ is the reduction in trap frequency.
The partition function $Z_{z,l}$ and weight $\zeta_{z,l}$ are also dependent on $l$, for a Boltzmann distribution one has $\zeta_{z,l} = \exp(-\hbar\omega_zr_l/(k_BT_z))$.

The trap frequency inhomogeneity between layers is illustrated in Fig.~\subfigref{sideband_vertical}{b} for realistic parameters.
A reduction in trap depth affects the sideband shape in two ways.
First, it simply rescales the sideband frequency.
For sufficiently distinct $r_l$'s, the unique sideband frequency per layer can be resolved.
Second, it effectively rescales $\rec$, also modifying the tail of the sideband spectrum: for a fixed $T_r$, shallower layers have shorter tails towards the carrier.

\subsubsection{Tweezer}\label{sec:sideband_tweezer}
We assume a tightly focused Gaussian laser beam with wavelength $\lambda=2\pi/k$ propagating along $z$.
Since the waist $w_0$ is small and comparable to $\lambda$, we need to consider the evolution of the beam size over distances comparable to the Rayleigh range $z_R=k w_0^2/2$.
As a result, the trapping potential is
\be
	V(x,y,z) = - \frac{V_t}{1+(z/z_R)^2}\exp\left(-\frac{2(x^2+y^2)}{w_0^2\left(1+(z/z_R)^2\right)}\right),
\ee
where $V_t$ is the tweezer's depth.
The minimum of this potential is at $x=y=z=0$, where it can be expanded to quartic order as
\be
	\frac{V(x,y,z)}{V_t} \simeq -1 + \frac{1}{z_R^2}z^2 + \frac{2}{w_0^2}r^2 - \frac{1}{z_R^4}z^4 - \frac{4}{w_0^2z_R^2}r^2z^2 - \frac{2}{w_0^4}r^4,
\ee
where $r=\sqrt{x^2+y^2}$.
After diagonalization and perturbation theory, the eigenenergies are given by Eq.~\eqref{eq:Equartic} with
\be\begin{split}
	&\omega_x = \omega_y = \sqrt{4V_t/(mw_0^2)}=\frac{2\sqrt{2}\rec}{kw_0} \sqrt{\mathcal{V}_t} \equiv\omega_r,\\
	&\omega_z=\sqrt{2V_t/(mz_R^2)}=\omega_r\sqrt{2}/(kw_0),\\
	&\Omega_{xx}=\Omega_{yy}=-3\rec\omega_z^2/(4\omega_r^2),\\
	&\Omega_{zz} = -3\rec\omega_z^4/(2\omega_r^4),\\
	&\Omega_{xz}=\Omega_{yz}=-2\rec\omega_z^3/\omega_r^3,\\
	&\Omega_{xy}=-\rec\omega_z^2/\omega_r^2,
\end{split}\ee
where $\mathcal{V}_t = V_t / (\hbar\rec)$.
If the probe beam is oriented along $z$, as we have considered thus far, we can again take advantage of the symmetry of the potential and treat $x$ and $y$ on equal footing, yielding a blue sideband frequency of
\be
	\tilde\omega_{+1}(n_r,n_z)=\omega_z - 3\rec\frac{\omega_z^4}{\omega_r^4}(n_z+1) - 2\rec\frac{\omega_z^3}{\omega_r^3}(n_r+1).
\ee
Since the numerical aperture is limited to 1, we generically have $\omega_z/\omega_r<1$.
Hence, we can again eliminate $n_r$.
Solving $\tilde\omega = \tilde\omega_{+1}(n_r,n_z)$ for $n_r$ yields
\be\begin{split}
	\tilde{n}_r(\tilde\omega)&=\frac{\omega_r^3}{2\rec\omega_z^3}\left(\omega_z-\tilde\omega-3\rec\frac{\omega_z^4}{\omega_r^4}(n_z+1)\right) - 1\\
&\equiv \frac{\omega_r^3}{2\rec\omega_z^3}s_{t,z}(\tilde\omega; n_z) - 1,
\end{split}\ee
where $s_{t,z}(\tilde\omega; n_z)$ is different from $s_z(\tilde\omega; n_z)$ in its dependence on $n_z$.
The approximated sideband kernel follows from plugging this into Eq.~\eqref{eq:kernel_radial}, and we find a result given by Eqs.~\eqref{eq:kernel1d} and \eqref{eq:kernel1d_fun}, though with $s_z(\tilde\omega; n_z)$ replaced by $s_{t,z}(\tilde\omega; n_z)$ above, and with $\alpha$ replaced by $\alpha_t=\beta_r\omega_r^3/2\rec\omega_z^3$.
The sideband shape after Lorentzian convolution is then given by Eq.~\eqref{eq:sideband1d} with the same modifications applied.
The intuition discussed there about the relation between sideband spectrum and temperature thus still holds, though we note that now the sideband shape depends on $\omega_r$ as its dependence does not cancel out inside $\alpha_t$.

Next, let us consider a probe beam propagating along one of the radial directions, choosing $x$ w.l.o.g.
The general framework still applies in this situation, but the blue sideband is now given by $\Delta\mathbf{n}=(+1,0,0)$, which in this case becomes
\be
	\tilde\omega_{+1}(n_x,n_y,n_z) = \omega_r - \frac{\rec\omega_z^2}{2\omega_r^2}(3n_x + 2n_y + 4) - \rec\frac{\omega_z^3}{\omega_r^3}(2n_z+1).
\ee
The weakest dependence is on $n_z$ which we will eliminate via
\be\begin{split}
	\tilde{n}_z(\tilde\omega) &= \frac{\omega_r^3}{2\rec\omega_z^3}\left(\omega_r - \tilde\omega - \rec\frac{\omega_z^2}{2\omega_r^2}(3n_x+2n_y+4)\right) - \frac{1}{2}\\
& \equiv \frac{\omega_r^3}{2\rec\omega_z^3}s_{t,r}(\tilde\omega;n_x,n_y) - \frac{1}{2}.
\end{split}\ee
We then have
\be
	\rho_{+1}(\tilde\omega)\approx\frac{\omega_r^3}{2\rec\omega_z^3}\sum\limits_{n_x,n_y=0}^N f_T(n_x,n_y,\tilde{n}_z(\tilde\omega))\Theta(\tilde{n}_z(\tilde\omega)).
\ee
Note that since the probe is along $x$, the modified sideband kernel $\rho_{(n_z^\text{min})}(\tilde\omega)$ only adjusts the lower bound of the summation over $n_x$.
Since $n_x$ and $n_y$ appear separately in $\tilde{n}_z(\tilde\omega)$, i.e.\ not in the particular combination $n_x+n_y$, we unfortunately cannot reduce it to a single summation this way.
However, we note that the dependence on $n_y$ is only subtle.
First, the presence of $\tilde{n}_z(\tilde\omega)$ in the Boltzmann factor slightly `renormalizes' the temperature for $y$, but proper adjusting of the partition function $Z_y$ counteracts this.
The only other dependence is in the edge of the step function $\Theta$, but there absolute shifts of less than $\Gamma$ would be washed out.
Considering the tweezer provides tight confinement along $y$, the upper limit for the summation over $n_y$ is relatively small, and we recall that $\omega_z/\omega_r<1$.
We thus further approximate the sideband kernel by ignoring this dependence.
A comparison between the end result of this approximation and the exact result given by Eq.~\eqref{eq:sideband_exact} shows good correspondence, as we have numerically verified.

\begin{figure}[t!]
	\includegraphics{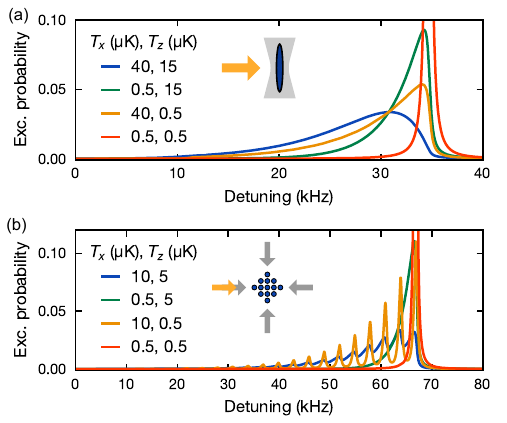}
	\caption{\label{fig:sideband_2d}
		\textbf{Sideband shape in 2D geometries}.
        \panel{a,b} Illustration of sideband shapes for different temperatures.
        Both use $\Gamma=2\pi\times0.5\kHz$ and $\rec=2\pi\times3\kHz$; the red curve is cropped for visibility reasons, which has a single, sharp peak.
		\panel{a} Radial probing of a tweezer (see inset) yields a sideband given by Eq.~\eqref{eq:sideband_tweezer}.
        For all curves we take $\omega_r = 2\pi\times35\kHz$, $\omega_z = 2\pi\times8\kHz$, corresponding to $w_0=\lambda$ and a depth of $\mathcal{V}_t\simeq670$ ($\simeq100\uK$).
        The sideband spectra are more sharply peaked than in 1D, cf.~Fig.~\ref{fig:sideband_1d}.
		\panel{b} On-axis probing of a homogeneous crossed 2D lattice yields a sideband given by Eq.~\eqref{eq:sideband_centraltube}.
        We use $\omega_x=2\pi\times70\kHz$ and $\omega_z=2\pi\times0.3\kHz$.
        Longitudinal anharmonicity is more severe than for tweezers, resulting in larger, clearly resolvable peaks.
        The inset schematically illustrates the probe beam (yellow arrow) orientation compared to the lattices (gray arrows).}
\end{figure}

With this further approximation, the summation over $n_y$ is trivial thanks to the normalization of Boltzmann factors, and we arrive at
\begin{align}
	&\rho_{+1}(\tilde\omega)\approx\frac{C'}{Z_x}\sum\limits_{n_x=0}^N \zeta_x^{n_x}f_{2\text{D}}\left(s_{t,r}(\tilde\omega; n_x, 0)\right),\\
    &f_{2\text{D}}(s) = e^{-\beta s}\Theta(s),
\end{align}
with $Z_x = \frac{1-\zeta_x^{N+1}}{1-\zeta_x}$, $\beta = \beta_z\omega_r^3/(2\rec\omega_z^3)$, where we ignored a small contribution inside the step function as before, and $C'=\beta$ ensures area-normalization.
As with the 1D sideband, the convolved sideband can be found by analytically evaluating the Lorentzian convolution for $f_{2\text{D}}(s)$, which gives
\be\label{eq:conv_2d}
	I_{2\text{D}}(y) = \int\limits_{-\infty}^\infty \text{d}x f_{2\text{D}}(x) L_\Gamma(y - x) = \frac{\Gamma}{4}\Im\{e^{-z} \Gamma(0,-z) \},
\ee
with $z=\beta(y+i\Gamma/2)$.
The blue sideband shape is then
\be\label{eq:sideband_tweezer}
	P_{+1}(\tilde\omega) = \frac{\beta}{Z_x}\sum\limits_{n_x=0}^N\zeta_x^{n_x}I_{2\text{D}}(s_{t,r}(\tilde\omega; n_x, 0)).
\ee
Compared to the 1D sideband shape, this sideband is markedly sharper, as can be seen by comparing the functional forms of $f_{2\text{D}}(s)$ and $f_{1\text{D}}(s)$.
This difference stems from the absence of a radial density of states: in a 2D trap such as a tweezer there is only one weakly confined direction, while in a 1D lattice there are two.
The sharper peak is also prominent in Fig.~\subfigref{sideband_2d}{a}, where the sideband shape is illustrated for different temperatures.
Additionally, through these approximations, the sideband shape does not depend on the population distribution for $n_y$ and hence $T_y$; recall that $y$ is the radial direction orthogonal to probing.
The width of the sideband is thus governed by $T_z$ and $T_x$, where we further note that it is hard to reach the regime where different $n_x \rightarrow n_x+1$ transitions are resolvable since the spacing is $3\omega_z^2\rec/(2\omega_r^2)$ as compared to $\rec/2$ in 1D.

\subsubsection{Crossed 2D lattice}
We now consider the situation of two independent 1D retro-reflected beams with wavelength $\lambda$ that cross each other and (locally) create a 2D lattice.
We assume they have sufficiently distinct frequencies such that cross-interference can be neglected, which is typically the case for frequency differences $>20\MHz$.
In order to compare results with the radial tweezer spectrum discussed in Sec.~\ref{sec:sideband_tweezer}, we assume one beam propagates along $x$ and the other along $y$, and the probe is along $x$.
Each beam then creates an independent potential, with functional form as defined in Eq.~\eqref{eq:lattice1d}, so that the total potential is
\be\begin{split}
	V(x,y,z) = &- V_x\cos^2(kx)\exp\left(-2(y^2+z^2)/w_0^2\right)\\
    &- V_y\cos^2(ky)\exp\left(-2(x^2+z^2)/w_0^2\right).
\end{split}\ee
Here we assumed both beams have identical waist, but note that generalization to different waists is straightforward.
This combined potential generates a grid of tubes, with trap minima at $z=0$ and $x=l\lambda/2$, $y=m\lambda/2$ for $l,m\in\mathbb{Z}$.
(The Gaussian envelope marginally shifts the minima away from this location for large $\abs{l}$ or $\abs{m}$ but we will ignore this effect, see also the discussion below.)
Since the total potential is not translationally invariant along $x$ or $y$, we must analyze each minimum separately.
As we will see, the local trap shape varies across tubes, causing inhomogeneous broadening of the sideband shape.

We start our analysis by focusing on the central tube, i.e., the one at $x=0$, $y=0$.
Here we locally expand the trap potential as
\be\begin{split}
	V&(x,y,z) \simeq - (V_x+V_y) \\
    &+ \left(k^2V_x + \frac{2V_y}{w_0^2}\right)x^2 + \left(\frac{2V_x}{w_0^2} + k^2V_y\right)y^2 + \frac{2(V_x+V_y)}{w_0^2}z^2 \\
    & - \left(\frac{1}{3}k^4V_x + \frac{2V_y}{w_0^4}\right)x^4 - \frac{2k^2(V_x+V_y)}{w_0^2}x^2y^2\\
    & - \frac{2}{w_0^2}\left(k^2V_x+\frac{2V_y}{w_0^2}\right)x^2z^2 + \ldots,
\end{split}\ee
where we omitted the $y^4$, $z^4$, and $y^2z^2$ terms since they are irrelevant in the sideband frequency.
The resulting eigenenergies are obtained using the following expressions
\be\begin{split}
	&\omega_x = 2\rec\sqrt{\mathcal{V}_x},\\
	&\omega_y = 2\rec\sqrt{ \mathcal{V}_y},\\
	&\omega_z = \sqrt{4(V_x+V_y)/(m w_0^2)},\\
	&\Omega_{xx}=\Omega_{yy}=-\frac{1}{2}\rec,\\
	&\Omega_{zz} =-\frac{3\rec}{2k^2w_0^2},\\
	&\Omega_{xy}=-\rec\omega_z^2/(\omega_x\omega_y),\\
	&\Omega_{xz}=-2\rec\omega_x/(\omega_z k^2w_0^2),\\
	&\Omega_{yz}=-2\rec\omega_y/(\omega_z k^2w_0^2),
\end{split}\ee
where $\mathcal{V}_i = V_i/(\hbar\rec)$ and again we have suppressed subleading orders in $1/(kw_0)$ for notational convenience.
The blue sideband frequency is thus given by
\be\begin{split}
	\tilde\omega_{+1}&(n_x,n_y,n_z) = \omega_x - \rec\left(n_x+1\right) \\
    &-\frac{\rec\omega_z^2}{2\omega_x\omega_y}(2n_y+1) - \frac{\rec\omega_x}{\omega_z k^2 w_0^2}(2n_z+1).
\end{split}\ee
Next, we need to decide which direction to eliminate.
The prefactors of $n_y$ and $n_z$ are $\abs{\partial\tilde\omega_{+1}/\partial n_y}=\rec\omega_z^2/(\omega_x\omega_y)$ and $\abs{\partial\tilde\omega_{+1}/\partial n_z}=2\rec\omega_x/(\omega_zk^2w_0^2)$.
Both these sensitivities are small, because $\omega_z/\omega_{x,y}=O(k^{-1}w_0^{-1})\ll1$ and typically $\rec/\Gamma\lesssim1$.
However, the summation over $n_y$ runs only to $N_y\sim V_y/(\hbar\omega_y)$ which is much smaller than that for $n_z$, $N_z\sim (V_x+V_y)/(\hbar\omega_z)$, again because $\omega_z\ll\omega_{x,y}$.
We thus argue that approximating the summation over $n_z$ with its Riemann integral provides a more accurate result than doing the same for $n_y$, which is confirmed by numerical verification.
To eliminate $n_z$ we find
\be\begin{split}
	\tilde{n}_z(\tilde\omega) + \frac{1}{2} = \frac{\omega_zk^2w_0^2}{2\rec\omega_x}\bigg(\omega_x& - \tilde\omega -\rec(n_x+1) \\
    &- \frac{\rec\omega_z^2}{2\omega_x\omega_y}(2n_y+1)\bigg).
\end{split}\ee
The dependence of $\tilde{n}_z(\tilde\omega)$ on $n_y$, captured by the second line, is very weak.
As with the tweezer potential, this dependence renormalizes the Boltzmann factor for $y$ and shifts the edge of the step function $\Theta$.
The first effect does not affect the sideband shape.
For the latter effect, the relevant scale is $N_y\rec\omega_z^2/(\omega_x\omega_y)=N_y\abs{\partial\tilde\omega_{+1}/\partial n_y}$, which needs to be compared to $\Gamma$.
We already saw above that $\abs{\partial\tilde\omega_{+1}/\partial n_y}\ll\Gamma$, and the upper bound of the summation over $n_y$ is typically $O(10)$.
As a result, the shifting of the sideband edge is negligible, and we can ignore the dependence on $n_y$ entirely.
In other words, we approximate
\be\begin{split}
	\tilde{n}_z(\tilde\omega) &\approx \frac{\omega_zk^2w_0^2}{2\rec\omega_x}\left(\omega_x - \tilde\omega -\rec(n_x+1)\right) \\
    &\equiv \frac{\omega_zk^2w_0^2}{2\rec\omega_x}s_x(\tilde\omega; n_x),
\end{split}\ee
where we also omitted the shift by $1/2$ as we did before.
We can now substitute this into Eq.~\eqref{eq:rho_integral}.
The summation over $n_y$ is trivial since the summand no longer depends on $n_y$, resulting in
\be
	\rho_{+1}(\tilde\omega) \approx \frac{C'}{Z_x} \sum\limits_{n_x=0}^N e^{-\beta s_x(\tilde\omega; n_x)}\zeta_x^{n_x}\Theta(s_x(\tilde\omega; n_x)),
\ee
where $Z_x = \frac{1-\zeta_x^{N+1}}{1-\zeta_x}$, $\beta = \beta_z\omega_zk^2w_0^2/(2\rec\omega_x)$, and $C'=\beta$ again ensures normalization.
We recognize the same functional form
\be
    f_{2\text{D}}(s) = e^{-\beta s}\Theta(s)
\ee
as for the radial tweezer spectroscopy.
This is not surprising: the basic shape of the trapping potential of the 2D lattice is not so different compared to that of a tweezer, i.e., cigar-shaped.
Of course, anharmonicities and their cross-couplings are different, resulting in the modified expression for $\beta$.
Then, the result of the Lorentzian convolution is again given by Eq.~\eqref{eq:conv_2d}, and
\be\label{eq:sideband_centraltube}
	P_{+1}(\tilde\omega) = \frac{\beta}{Z_x}\sum\limits_{n_x=0}^N\zeta_x^{n_x}I_{2\text{D}}(s_x(\tilde\omega; n_x)),
\ee
with $s_x(\tilde\omega; n_x)$ as defined above.
This function's shape for typical parameters is illustrated in Fig.~\subfigref{sideband_2d}{b}.
Compared to the sideband generated by tweezers, shown in Fig.~\subfigref{sideband_2d}{a}, the most notable difference is the reappearance of a regime (high $T_x$, and $T_z$ not too high) where the longitudinal modes are resolved, reminiscent of that seen in 1D (recall Fig.~\ref{fig:sideband_1d}).

\begin{figure*}[t!]
	\includegraphics{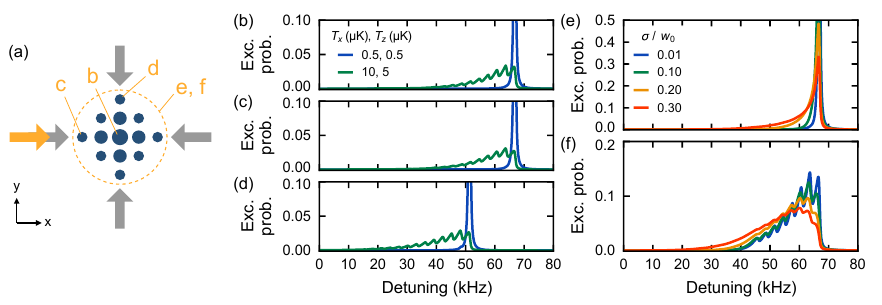}
    \caption{\label{fig:sideband_2d_inhom}
	\textbf{Inhomogeneity in a 2D lattice}.
    \panel{a} Schematic illustration of trap depth inhomogeneity across tubes of a 2D lattice, resulting in inhomogeneous broadening of sideband spectra.
    Annotations are the legend for the sideband spectra in panels \panel{b-f}, showing either individual tubes [panels \panel{b-d}] or the ensemble-average [panels \panel{e,f}].
    \panel{b} Sideband shape for the central tube, reproduced from Fig.~\subfigref{sideband_2d}{b}.
    Trap potential parameters are identical to that figure.
    \panel{c} Sideband shape for a tube displaced along the direction of probing, by $\Delta x = 0.5w_0$, resulting in negligible difference.
    \panel{d} Sideband shape for a tube displaced transversally to the probe, by $\Delta y = 0.5w_0$.
    The reduction in longitudinal trap frequency shifts the sideband closer to zero detuning, but all other features remain.
    \panel{e,f} Ensemble-averaged sideband shapes depending on the atomic cloud size $\sigma$ compared to the lattice beam waist $w_0$.
    Trapping parameters are consistent with previous panels, requiring $w_0/\lambda\simeq 75$, but sideband linewidth is now related to a carrier Rabi frequency of $\Omega = 2\pi\times2\kHz$.
    Temperatures correspond to panels \panel{b-d}, i.e., $(T_x,T_z)=(0.5, 0.5)\uK$ for panel \panel{e} and $(T_x,T_z)=(10, 5)\uK$ for panel \panel{f}.
    The ensemble-average is evaluated using Eq.~\eqref{eq:sideband2d_cheap} and coarse-graining parameter $B=50$.}
\end{figure*}

The analysis presented above applies to the central tube located at $x=0$, $y=0$.
An atomic ensemble loaded into this lattice potential will generally also populate other tubes.
The other tubes are located at minima of the potential, i.e.\ where $\vec{\nabla} V = \vec{0}$.
Since $w_0 \gg \lambda$, the Gaussian envelope of each potential can be considered a slowly evolving function, and the minima are located at $x=l\lambda/2$ and $y=m\lambda/2$.
There, locally the potential shape is similar to that at $x=0$, $y=0$ except with the renormalization $V_x \rightarrow V_x(m) = V_x e^{-m^2\lambda^2/(2w_0^2)}$ and $V_y \rightarrow V_y(l) = V_y e^{-l^2\lambda^2/(2w_0^2)}$.
The sideband spectrum will be an inhomogeneous average over all tubes, given by
\be\label{eq:sideband2d}
	\overline{P}_{+1}(\tilde\omega) = \sum\limits_{l,m} p(l,m) P_{+1}\left(\tilde\omega; V_x(m), V_y(l)\right),
\ee
where $p(l,m)$ is the population fraction of the atomic ensemble that resides in the tube indexed by $l$ and $m$, and we made explicit the dependence of $P_{+1}$ on the lattice depths.
This average can be calculated either for the exact spectrum given by Eq.~\eqref{eq:sideband_exact}, or using the analytical approach.
The latter has an additional benefit of clarifying the dependence on $V_x$ and $V_y$.
Since non-central tubes solely have rescaled depths $V_x(m)$ and $V_y(l)$, their sideband shape is still given by Eq.~\eqref{eq:sideband_centraltube}, but with modified parameters.
As we can see in the preceding expressions, $\beta$ and $s_x(\tilde\omega;n_x)$ do not depend on $V_y$ to leading order, leaving the only dependence of the sideband shape on $V_y$ through $\omega_{z}$.
Nonetheless, this has a rather subdued effect, as illustrated in Fig.~\ref{fig:sideband_2d_inhom}, where the sideband shape of atoms in the central tube [Fig.~\subfigref{sideband_2d_inhom}{b}] and a tube displaced along the direction of probing [Fig.~\subfigref{sideband_2d_inhom}{c}] appear practically identical.
A much stronger response is elicited when comparing to a tube displaced orthogonal to the probe beam [Fig.~\subfigref{sideband_2d_inhom}{d}], as that probes the dependence on $V_x$ which is evident.
We can thus approximate Eq.~\eqref{eq:sideband2d} as
\be\label{eq:sideband2d_cheap}
	\overline{P}_{+1}(\tilde\omega) \approx \sum\limits_{m}p(m)P_{+1}\left(\tilde\omega; V_x(m), V_y(0)\right),
\ee
where $p(m)$ is the marginal of $p(l,m)$.
For practical evaluation of Eq.~\eqref{eq:sideband2d_cheap} (or Eq.~\eqref{eq:sideband2d} above), we can utilize that $V_x(m)$ (and $V_y(l)$) evolve(s) smoothly over a scale set by $w_0$ and $\lambda\ll w_0$.
Thus, we have $V_x(m+1)\approx V_x(m)$ and it is unnecessary to evaluate the ensemble average tube-by-tube and we can instead use coarse-graining.
The relevant scale for $l$ and $m$ is set by the extent of the atomic cloud, i.e.\ the widths of $p(l,m)$, provided it is smaller than $w_0$.
Assuming an isotropic cloud with a Gaussian density distribution of standard distribution $\sigma$, the coarse-graining ``block size'' can be chosen as $\sigma/B$, where $B$ controls the precision.
Figures~\subfigref{sideband_2d_inhom}{e,f} show the effect of this inhomogeneous broadening on the shape of the sideband for different cloud sizes $\sigma$, for two different temperature regimes.
The resultant shape retains features seen in the sideband spectrum of the central tube, but clearly exhibits a wider shape due to the non-negligible extent of the atomic cloud.
This effect is especially noticeable when the sideband width is otherwise narrow, i.e.\ when the atoms are in the motional ground state.

In closing this Subsection we reiterate that these results hold for a 2D lattice made of two independent beams.
For the case of a folded 2D lattice, where the two crossing arms are at the same optical frequency and polarization and thus interfere, most considerations will still apply but precise expressions for e.g.,\ $s_x(\tilde\omega;n_x)$ and $\beta$ may be different due to the modified local geometry of the trapping potential.

\subsubsection{3D lattice}
Finally, we discuss sideband spectroscopy in a 3D lattice.
We start by considering a 3D lattice formed by three independent 1D retro-reflected beams, akin to the 2D lattice above.
Once again ignoring interference between the beams as well as their Rayleigh range, the total potential is
\be \begin{split}
	V(x,y,z) = &- V_x\cos^2(kx)\exp\left(-2(y^2+z^2)/w_0^2\right)\\
    &- V_y\cos^2(ky)\exp\left(-2(x^2+z^2)/w_0^2\right)\\
    &- V_z\cos^2(kz)\exp\left(-2(x^2+y^2)/w_0^2\right).
\end{split}\ee
This results in a 3D grid of trap minima, which we refer to as the lattice sites, at $x=l\lambda/2$, $y=m\lambda/2$, $z=n\lambda/2$ for $l,m,n\in\mathbb{Z}$.
As before, we need to carefully consider site-to-site inhomogeneity, but we first consider the central lattice site at $(x,y,z) = (0,0,0)$.
Expanding the trap potential as before, and ignoring subleading contributions, results in eigenenergies with
\be\begin{split}
	&\omega_x = 2\rec\sqrt{\mathcal{V}_x},\\
	&\omega_y = 2\rec\sqrt{\mathcal{V}_y},\\
	&\omega_z = 2\rec\sqrt{\mathcal{V}_z},\\
	&\Omega_{xx} = \Omega_{yy} = \Omega_{zz} = -\frac{1}{2}\rec,\\
	&\Omega_{xy}= -2\rec(\omega_x^2 + \omega_y^2)/(\omega_x\omega_y k^2w_0^2),\\
	&\Omega_{xz}= -2\rec(\omega_x^2 + \omega_z^2)/(\omega_x\omega_z k^2w_0^2),\\
	&\Omega_{yz}= -2\rec(\omega_y^2 + \omega_z^2)/(\omega_y\omega_z k^2w_0^2).
\end{split}\ee
For probing along $z$ the sideband frequency is thus
\be\begin{split}
	\tilde\omega_{+1}&(n_x,n_y,n_z)\approx \omega_z - \rec(n_z+1)\\
    &- \frac{\rec}{k^2w_0^2}\left(\frac{\omega_x^2 + \omega_z^2}{\omega_x\omega_z}(2n_x+1) + \frac{\omega_y^2 + \omega_z^2}{\omega_y\omega_z}(2n_y+1)\right),
\end{split}\ee
and the result for probing in other directions follows from appropriate permutation of indices.
The prefactors within the parentheses in the second line are both $O(1)$, provided all three lattice beams are of roughly comparable strength.
Assuming $k^{-2}w_0^{-2}\ll 1$ and $\rec \lesssim \Gamma$, the last term then has no significant effect on the sideband shape.
Given this observation, we choose to ignore the dependence of $\tilde\omega_{+1}$ on $n_x$ and $n_y$ altogether, rather than eliminate summations over these indices according to the procedure outlined in Section~\ref{sec:sideband-framework}.
Then, the sideband kernel is simply
\be\label{eq:kernel_centralsite}
	\rho_{+1}(\tilde\omega) = \sum\limits_{n_z=0}^N f_T(n_z)\cdot\delta\left(\tilde\omega - \tilde\omega_{+1}(n_z)\right).
\ee
The convolution with the Lorentzian then also is particularly simple, giving
\be\label{eq:sideband_centralsite}
	P_{+1}(\tilde\omega) = \frac{1}{Z_z}\sum\limits_{n_z=0}^N\zeta_z^{n_z} L_\Gamma\left(\omega_z - \tilde\omega - \rec(n_z+1)\right).
\ee
Summarizing, in a 3D lattice confinement is strong enough that atoms are unable to sample regions of the potential where cross-coupling between the different axes is evident.
Thus, the sideband spectrum shape depends solely on the local longitudinal trap frequency and temperature, the latter affecting the degree to which atoms can experience anharmonicity of the potential in the longitudinal direction.

\begin{figure}[t!]
	\includegraphics{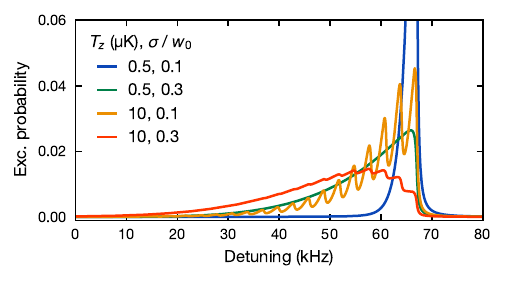}
	\caption{\label{fig:sideband_3d}
		\textbf{Sideband shape in a 3D lattice}.
        Dependence of sideband shape on the longitudinal temperature and the extent of the atomic cloud.
        In a 3D lattice, the sideband shape does not depend on anything else, see text for discussion.
        For all curves we take $\omega_z = 2\pi\times70\kHz$, $\rec = 2\pi\times3\kHz$, and $\Gamma=2\pi\times0.5\kHz$.
        Inhomogeneous broadening is evident, akin to that seen in a 2D lattice.}
\end{figure}

Of course, we still need to consider the inhomogeneous average over all lattice sites.
This is similar to the 2D lattice, where now we have to renormalize all three potential strengths based on $x$, $y$, and $z$.
However, since only $\omega_z$ appears in Eq.~\eqref{eq:sideband_centralsite}, only the renormalization of $V_z$ affects the spectrum, allowing us to write
\be\label{eq:sideband3d}
	\overline{P}_{+1}(\tilde\omega) = \sum\limits_{l,m} p(l,m)P_{+1}\left(\tilde\omega; V_z(l,m)\right),
\ee
where $V_z(l,m) = V_z \exp(-(l^2+m^2)\lambda^2/(2w_0^2))$ and $p(l,m)$ is the population distribution of the atomic ensemble in the two directions orthogonal to the probe beam (i.e., $x$ and $y$).
This time, instead of coarse-graining, we can argue that $V_z(l,m)$ evolves smoothly with respect to $l$ and $m$, allowing us to approximate the summation averaging over lattice sites by its Riemann integral, yielding
\be
	\overline{P}_{+1}(\tilde\omega) \approx \int \text{d}x \text{d}y \, p(x,y) P_{+1}\left(\tilde\omega; V_z(x,y)\right),
\ee
where we went back to integrating in real space rather than indices.
We can swap the order of operations between the summation over $n_z$ inside $P_{+1}(\tilde\omega)$ and the integral, and perform the resulting integral analytically.
Doing this first for the sideband kernel we find
\be\label{eq:kernel3d}
	\overline{\rho}_{+1}(\tilde\omega) \approx \frac{1}{Z_z}\sum\limits_{n_z=0}^N\zeta_z^{n_z}\overline{f_{3\text{D}}}\left(\tilde\omega + \rec(n_z+1)\right),
\ee
with
\be\begin{split}
	\overline{f_{3\text{D}}}(s) &= \int\limits_{-\infty}^\infty \text{d}x\int\limits_{-\infty}^\infty \text{d}y\,p(x,y) \delta\left(s - \omega_z e^{-(x^2+y^2)/w_0^2}\right)\\
	&=\sigma^{-2}\int\limits_0^\infty r \text{d}r\,e^{-r^2/(2\sigma^2)} \delta\left(s - \omega_z e^{-r^2/w_0^2}\right)\\
	&=\frac{\gamma}{\omega_z} \left(\frac{s}{\omega_z}\right)^{\gamma-1}\Theta(\omega_z - s)\Theta(s),
\end{split}\ee
where we assumed an isotropic Gaussian (normal) population distribution with standard deviation $\sigma$, used the compounded sifting property of the Dirac delta-function, $\gamma = w_0^2/(2\sigma^2)$, and $\omega_z$ is the trap frequency of the central lattice site, i.e.\ $\omega_z\approx 2\rec\sqrt{\mathcal{V}_z}$.
The site-averaged sideband shape is then simply the convolution $\overline{P}_{+1}(\tilde\omega) = L_\Gamma(\tilde\omega) * \overline{\rho}_{+1}(\tilde\omega)$, for which we only need
\be\begin{split}
	I_{3\text{D}}(s) &\equiv L_\Gamma(s) * \overline{f_{3\text{D}}}(s)\\
    &= L_\Gamma(s) \Re\left\{\frac{2i\upsilon^*}{\Gamma}\pFq{2}{1}\left(1,\gamma, 1+\gamma, \frac{\omega_z}{\upsilon}\right)\right\},
\end{split}\ee
where $\Re\{\cdot\}$ denotes the real part, $\pFq{2}{1}$ is the (ordinary) hypergeometric function~\cite{Abramowitz_Handbook_1972} and $\upsilon = s + i\Gamma/2$.
Evidently, the sideband shape depends largely on the cloud size $\sigma$ compared to the beam waist $w_0$, with the only other dependence being on the longitudinal temperature $T_z$.
The effect of this is illustrated in Fig.~\ref{fig:sideband_3d}.
Finally we note that spectroscopic probing along other principal directions of the 3D lattice results in identical expressions except with the corresponding substitution of the coordinate label, e.g.\ $\omega_z\rightarrow\omega_x$, etc.

Above, we considered a 3D lattice with $\lambda/2$ spacing between lattice sites in all directions.
By virtue of our lattice along $z$ being a shallow-angle interference lattice, this is not true in that direction, resulting in some modifications as we saw in Section~\ref{sec:sideband_shallow-angle}.
This also breaks the symmetry between the three probing directions and further complicates the inhomogeneous average across lattice sites, as we now describe.
First, as long as the confinement is sufficient in all three directions, the sideband of any individual lattice site is not modified and remains solely dependent on the trap frequency and occupation number along the direction of probing, e.g.\ $\omega_z$ and $n_z$ for probing along $z$, as in Eqs.~\eqref{eq:kernel_centralsite} and \eqref{eq:sideband_centralsite}.
However, we need to revisit the inhomogeneous average across lattice sites.

For probing along $z$, within each layer all expressions above hold except that we need to use the layer-specific renormalizations of trap frequency and recoil energy as in Sec.~\ref{sec:sideband_shallow-angle}.
Explicitly, the kernel averaged over all sites is
\be
    \overline{\rho}_{+1}(\tilde\omega)=\sum\limits_n\frac{p_n}{Z_{z,n}}\sum\limits_{n_z=0}^N\zeta_{z,n}^{n_z}\overline{f_{3\text{D}, n}}\left(\tilde\omega + \rec r_n^2\sin^2(\theta)(n_z+1)\right),
\ee
where the layers are indexed by $n$; recall that $p_n$ is the layer's population and that $r_n = \exp(-\lambda^2(n+\frac{\phi}{2\pi})^2/(4\tan^2(\theta)w_V^2))$, and we have
\be
    \overline{f_{3\text{D}, n}}(s) = \frac{\gamma_n}{\omega_zr_n} \left(\frac{s}{\omega_zr_n}\right)^{\gamma_n-1}\Theta(\omega_zr_n - s)\Theta(s),
\ee
where we also allow $\gamma$ to vary between layers.

For probing along the horizontal directions, choosing again $x$, the sideband spectrum is given by Eq.~\eqref{eq:sideband3d} but with summation over $m$ and $n$ and dependence on $V_x(m,n)$.
Since the number of vertical layers is limited, the summation along that direction does not lend itself to approximation with its Riemann integral.
Rather, we will integrate the remaining horizontal direction $y$ and simply sum the layers directly.
This yields
\be
    \overline{\rho}_{+1}(\tilde\omega) = \sum\limits_n\frac{p_n}{Z_x}\sum\limits_{n_x=0}^N\zeta_x^{n_x}\overline{f_{3\text{D}^*, n}}\left(\tilde\omega+\rec(n_x+1)\right),
\ee
where
\be\begin{split}
	\overline{f_{3\text{D}^*,n}}(s) &= \int\limits_{-\infty}^\infty \text{d}y\, p(y) \delta\left(s - \omega_x e^{-(y^2+z(n)^2)/w_0^2}\right) \\
	&= \sqrt{\frac{\gamma}{\pi}} \frac{1}{s\sqrt{-\log\left(\frac{s}{\omega_x(n)}\right)}} \left(\frac{s}{\omega_x(n)}\right)^\gamma \\
    &\hspace{3em}\times\Theta(\omega_x(n) - s) \Theta(s),
\end{split}\ee
where $n$ indexes the layers at position $z(n)$, we assumed a Gaussian population distribution along $y$ with standard deviation $\sigma$, and we defined the layer-local trap frequency $\omega_x(n) = \omega_x \exp(-z(n)^2/w_0^2)$.
It can be straight-forwardly verified that $\bar{\rho}_{+1}(\tilde\omega)$ is still normalized as desired, provided $\sum\limits_n p_n = 1$.
The convolution with the Lorentzian to get $\overline{P}_{+1}(\tilde\omega)$ no longer has a closed form and has to be performed numerically.

\end{document}